\documentclass[aps,twocolumn,prd,superscriptaddress,showpacs,letterpaper,nofootinbib]{revtex4}
\usepackage{amsmath,amssymb}
\usepackage{color}
\usepackage[pdftex]{graphicx}
\usepackage[colorlinks=true,linkcolor=blue]{hyperref}
\usepackage{mathptmx}
\usepackage{comment}

\makeatletter
\newcommand{\rmnum}[1]{\romannumeral #1}
\newcommand{\Rmnum}[1]{\expandafter\@slowromancap\romannumeral #1@}

\begin{document}

\long\def\symbolfootnote[#1]#2{\begingroup%
\def\thefootnote{\fnsymbol{footnote}}\footnote[#1]{#2}\endgroup}


\title{Realistic polarizing Sagnac topology with DC readout for  the Einstein Telescope}

\author{Mengyao Wang}

\author{Charlotte Bond}

\author{Daniel Brown}

\author{Frank  Br\"{u}ckner}

\author{Ludovico Carbone}

\author{Rebecca Palmer}

\author{Andreas Freise}

\affiliation{School of Physics and Astronomy, University of Birmingham, Edgbaston, Birmingham B15 2TT, United Kingdom}


\begin{abstract}
The Einstein Telescope (ET) is a proposed future gravitational wave
detector. Its design is original, using a triangular
orientation of three detectors and a xylophone
configuration, splitting each detector into one
high-frequency and one low-frequency system. In other
aspects the current design retains the dual-recycled Michelson
interferometer typical of current detectors, such as Advanced LIGO.
In this paper, we investigate the feasibility of replacing the
low-frequency part of the ET detectors with a Sagnac
interferometer. We show that a Sagnac interferometer, using realistic
optical parameters based on the ET design, could provide a similar level of
radiation pressure noise suppression without the need for a signal recycling mirror
and the extensive filter cavities. We consider the practical issues of a realistic,
power-recycled Sagnac, using linear arm cavities and polarizing
optics. In particular, we investigate the effects of nonperfect polarizing optics
and propose a new method for the generation of a
local oscillator field similar to the DC readout scheme
of current detectors.
\end{abstract}

\maketitle


\section{Introduction}
\label{sec:introduction}

The Einstein gravitational-wave Telescope (ET) is a proposed
third-generation gravitational wave (GW) observatory. Its aim is to
achieve a factor of 10 improvement in sensitivity with respect to the advanced
detectors, such as Advanced LIGO\,\cite{Harry10} and Advanced VIRGO
\,\cite{Accadia11}, over a
broad range of frequencies\,\cite{Sathyaprakash12}.
The current design of ET is based
on three nested detectors, each being composed of two Michelson
interferometers (xylophone design\,\cite{shoemaker01,Hild10}), one optimized for low
frequencies (ET-LF) and the other for high frequencies (ET-HF).
Both interferometers have
10\,km long arm cavities  and use a {\it dual-recycled} Michelson
configuration
(combining {\it power recycling} and
{\it signal recycling}).
The xylophone design has been proposed to optimize the sensitivity between
2-40\,Hz and 40\,Hz-10\,kHz independently, allowing the separation of
cryogenic optics from high power laser beams. ET-HF is mainly concerned
with the photon counting noise (shot noise) and employs high laser power.
ET-LF was designed particularly to minimize the low-frequency quantum noise
caused by quantum fluctuations of the
photon number (radiation pressure noise) by using a low optical power.
ET has been envisaged as an infrastructure that could host different
implementations of GW detectors over a long time and
therefore the design offers the flexibility to choose different topologies and
configurations other than the dual-recycled
Michelson interferometer.

Initial proposals for interferometric measurements for GW detection
favored the Michelson interferometer because
it naturally provides a differential length measurement between
perpendicular arms.  This maximizes the signal for a GW (of one
polarization and direction) so is ideally suited for GW detection. The
original purpose of the Sagnac
interferometer was to measure rotation rather than mirror 
displacement\,\cite{sagnac13}. In 1995, successful
experimental tests of a zero-area Sagnac demonstrated a different
mode of operation, in which it becomes insensitive
to rotation but sensitive to mirror motion\,\cite{sun96}. The
zero-area Sagnac interferometer was thus revealed as an alternative
approach with the potential of reaching the
required sensitivity to detect GWs. Further investigations into the
performance and technical limitations of a Sagnac interferometric GW
detector has been developed\,\cite{Petrovichev98, Beyersdorf99, Beyersdorf99_b}. A reduced
susceptibility to laser frequency fluctuations
and imbalances of mirror positions was identified\,\cite{Petrovichev98} and
experimentally verified with a heterodyne detection\,\cite{Beyersdorf99}.
However, this comes at the cost of
tighter tolerances of the imperfections (i.e., mirror surface distortions)
and misalignments (i.e., beam splitter tilt, mirror tilt) of optical components
compared to the
Michelson topology\,\cite{Petrovichev98,Beyersdorf99_b}. The use of a Sagnac also
implies challenges in the length control of the combination of arm
cavities and dual-recycling mirrors\,\cite{Mizuno97}. These drawbacks
along with the rapid development of advanced techniques for the
further enhancement of the Michelson performance previously prevented a more
serious consideration of the Sagnac topology for GW detection. 

After significant efforts in reducing classical noise sources in the
GW community\,\cite{Harry10}, the impact of quantum noise is now one of the most
important challenges in designing future GW observatories, in particular the
Einstein Telescope\,\cite{ET11}. To reduce the quantum noise over a
broad frequency band, the baseline of
the current ET design includes the injection of frequency-dependent
squeezed
vacuum into the dark port of both the LF and HF Michelson
interferometers. The frequency-dependent rotation of the squeezing
angle is facilitated by implementing two
long filter cavities
for each interferometer\,\cite{ET11}. Alternatively, in 2003 Chen\,\cite{Chen03}
first described the quantum-nondemolition properties of
a Sagnac-type interferometer as it functions as a speed meter rather
than a position meter, which can remove the low-frequency radiation pressure
noise without the need of two long filter cavities. A comprehensive quantum
noise analysis of a
practical large-scale Sagnac interferometer (using LIGO parameters)
has been carried out\,\cite{Chen03, Danilishin04}, including the
consideration of optical losses. In\,\cite{Chen03}, the application of
km-scale ring cavity and delay line schemes in a Sagnac interferometer
has been investigated. The studies revealed very promising quantum
noise characteristics at low frequencies (1\,Hz - 100\,Hz) with only
little susceptibility to optical losses. A variant scheme is shown in
\,\cite{Danilishin04, Chen11, Danilishin12} with minimal changes to current
existing detector configurations by using polarizing optical
components (polarization speed meter based on Michelson configuration has recently been investigated by Wade {\it et al.}\,\cite{Wade2012}).  It has been shown that such a configuration using squeezed
vacuum injection but without filter cavities
can reduce low-frequency quantum noise to a similarly low level as a
Michelson with
filter cavities\,\cite{Danilishin12}. Recently, a tabletop
experiment has demonstrated
quantum noise reduction by nonfiltered squeezing in a Sagnac
interferometer\,\cite{Eberle10}.

During the ET design study the Sagnac topology was investigated as an
alternative option for the low-frequency part of the xylophone (ET-LF)
\,\cite{et_mueller_ebhard09}, focusing on a Sagnac design with ring cavities in
the interferometer arms.
However, ring cavities with a long baseline imply a number
of challenges, such as elliptical spot sizes on some mirrors and
a larger coupling of small-angle scattering back into the main beam.
Because of the vast experience using a Michelson interferometer within
the GW community and the advancements in technology specifically based
on the Michelson, a Sagnac topology was not chosen for the original ET
design.

In this paper, we reconsider the Sagnac topology for the
realization of the ET-LF interferometers. Precisely, we
compare the originally proposed Michelson-type interferometer
with two auxiliary filter cavities to a Sagnac-type interferometer
without filter cavities. The latter promises a significant reduction of
complexity and cost for the construction of the ET observatory whilst
achieving a compelling quantum-noise limited sensitivity. In contrast to earlier
considerations that investigated interferometers with ring cavities,
we study the Sagnac configuration with polarizing
optical components and linear Fabry-P\'{e}rot arm cavities. Technical
concerns of this configuration including the influence of
realistic polarizing optics on the dark fringe output and the
null response of a Sagnac to static mirror displacement are solved in an
elegant way.
We describe a novel method for using the leakage due to the
polarizers' finite extinction ratio to create a
\textit{Local Oscillator} (LO) for an optical readout scheme similar to the DC
readout adopted by advanced GW detectors\,\cite{adligo_design, Ward2008,
Hild09b, Fricke2012}. This has,
to the best of our knowledge, never been considered before. We analyze
the quantum noise behavior of a realistic ET-LF Sagnac instrument
with imperfect polarizing optics and frequency-independent
squeezing. With only minor adjustments on the ET-LF parameters, we
show that a comparable quantum noise level is feasible without the use
of filter cavities and also the signal recycling mirror.

The paper is organized as follows: Section~\ref{sec:configurations}
discusses different practical realizations of a polarizing Sagnac
interferometer with linear arm cavities.
One realization is selected
for this study and justified. In Sec.\,\ref{sec:QN},
we compute the quantum noise of the selected polarizing Sagnac
interferometer with ET-LF parameters. For simplicity, we start with
the case of perfect polarizing optics and illustrate the input-output relation
by using block diagrams.
Since we propose to use a slightly increased laser power,
adjustments of the mirror thermal noise and the effect on the overall
noise level are discussed briefly.
We do not attempt to propose a full alternative design for ET.
However, we present a discussion on the noise projection for a
Sagnac-type
design targeting a comparable quantum noise budget as the planned ET-LF
Michelson. Simple scaling laws are applied to the other noise sources.
We then extend our quantum noise model by considering a
finite extinction ratio to account for realistic polarizing beam splitters.
The quantum
noise level is evaluated for different degrees of imperfections and
varying parameters of the leakage-enabled DC readout scheme before a
sensitivity comparison between a Sagnac and the current
ET-LF design is presented. Requirements and different approaches for
creating a practical LO for a DC readout scheme are discussed in more detail
in Sec.\,\ref{sec:DC_readout}.
We also investigate different approaches to select a homodyne
detection angle and
present a novel method to correctly choose it. In the end of the section, we model
the polarizing
Sagnac interferometer using \textsc{Finesse}\,\cite{Freise04, finesse_webpage}
and show
the possibilities to control different degrees of freedom (DoFs).
In Sec.\,\ref{sec:conclusion} we give a
comprehensive conclusion of our findings.

 \section{Polarizing Sagnac Interferometer} \label{sec:configurations}
\begin{figure*}[t]
\begin{center}
\includegraphics[scale=0.78]{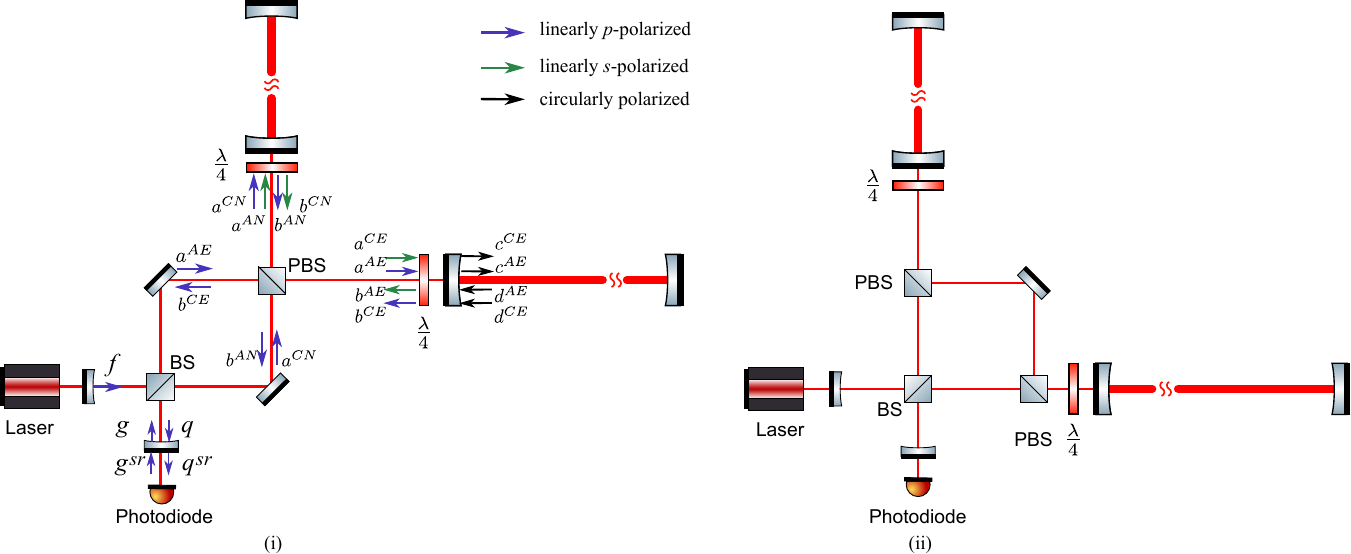}
\end{center}
\caption{Two versions of a polarizing Sagnac interferometer
  configuration. Both require minimum changes to the current ET-LF
  Michelson interferometers; relevant parameters are summarized in
  Table\,\ref{tab:configuration}. The interferometer has two linear 
  Fabry-P\'{e}rot arm cavities with a length of 10\,km. A PBS transmits
  \textit{p}-polarized and reflects
  \textit{s}-polarized light. A QWP transforms
  the linearly polarized beam into a circularly polarized beam and
  vice versa with a rotated 90$^{\circ}$ angle relative to the initial
  linear polarization. The left configuration which specifies each light field
  (i.e., a, b, \textit{etc}) will be used to evaluate the input-output relation in
  Sec.\,\ref{sec:PerfectPBSsag}. The arrows denote the {\it p}-polarized (blue)
  and {\it s}-polarized (green) beams which are respectively totally transmitted and
  reflected by an ideal PBS. Light fields circulating inside both arm cavities are
  circularly polarized (black) beams with both direction rotations.}
\label{fig:pbs_config}
\end{figure*}
 \begin{table}[b]
    \begin{center}
    \begin{tabular}{lcc}
    \hline
    \hline
    Parameter& Michelson & Sagnac \\
    \hline
    Arm length  & 10\,km & 10\,km   \\
    Distance PRM-BS & 10\,m  & 10\,m  \\
    Distance BS-ITM & 300 \,m & 300\,m\\
    Distance BS-PBS & ... &10\,m   \\
    Distance PBS-QWP & ... & 10\,m  \\
    Distance QWP-ITM & ... & 280\,m   \\
    \hline
    \end{tabular}
    \end{center}
    \caption{A table summarizing the baseline parameters of the ET-LF
      Michelson interferometer and an alternative Sagnac
      interferometer. The Michelson values are taken 
      from the original design study report~\cite{ET11}. The Sagnac values 
      are suggested by us for an ET Sagnac layout.}
    \label{tab:configuration}
    \end{table}
     
A realization of a Sagnac interferometer with minimum changes to the
planned ET infrastructure can be achieved by adding polarizing
beam splitters (PBSs) and quarter-wave plates (QWPs) to the current layout.
Two possible polarizing Sagnac configurations are shown in
Fig.\,\ref{fig:pbs_config}. They differ from a Michelson
interferometer as each beam after the central beam splitter (BS) travels through 
both arm cavities one after another. The
two beams share the same optical path but with opposite
propagation directions. This is achieved using PBSs and QWPs. The PBSs only 
transmit the \textit{p}-polarized
beam and reflect the \textit{s}-polarized.\footnote{We denote the component of the
electric field parallel to the {\it incident plane} as {\it p}-polarized and the
component perpendicular to the incident plane as {\it s}-polarized. Here, the
incident
plane is the plane made by the propagation direction and a vector perpendicular
to the reflection surface.} The
QWPs transform the linearly polarized beam into a circularly
polarized beam and again transform the circularly polarized beam into
linear polarization, rotated by 90$^{\circ}$ relative to the
initial linear polarization. Assuming ideal and lossless optical components, 
configurations (\rmnum{1}) and (\rmnum{2}) in Fig.\,\ref{fig:pbs_config} have the
same performance in
terms of quantum noise reduction. 
However, in reality each component will add optical losses and complexity.
The major difference between configuration (\rmnum{1}) and
(\rmnum{2}) is the number and position of the PBSs. 
Configuration (\rmnum{1}) has already been selected in previous investigations
for a polarizing Sagnac interferometer\,\cite{Danilishin04, Chen11, Danilishin12}.
An ET Sagnac topology baseline 
is given and the parameters with minimal and reasonable changes to the current 
Michelson baseline are summarized in Table
\ref{tab:configuration}. For the following discussion we have picked
configuration (\rmnum{1}) which features only one PBS and is thus
expected to exhibit lower optical losses due to finite extinction ratio.

\section{Quantum noise of a Sagnac Interferometer}\label{sec:QN}
We analyze the quantum noise behavior of a polarizing Sagnac
interferometer as shown in Fig.\,\ref{fig:pbs_config}\,(\rmnum{1}), taking into
account the imperfect nature of a realistic PBS.
To get the quantum noise spectral density (NSD), the input-output relation
of the interferometer needs to be specified \cite{Kimble02}.
Throughout this paper we present the input-output relations using
intuitive block diagrams. We first consider the case of an ideal PBS 
(Sec.\,\ref{sec:PerfectPBSsag}) and
then extend the model to include the effects of an imperfect PBS 
(Sec.\,\ref{sec:BadPBSsag}). In this paper,
we only consider the imperfection that a PBS has finite extinction ratio which 
induces a mixing of the two polarized fields. Optical
losses from arm cavities have been
included and investigated for both cases.
We compare the quantum noise for the same setup using PBSs with
different extinction ratios and propose an approach to increase the signal to
noise ratio of such a realistic Sagnac setup by using an additional PBS at the
detection port (to reduce the mixing of two polarized fields).

\subsection{Polarizing Sagnac interferometer with perfect PBS}
\label{sec:PerfectPBSsag}
\begin{figure}[t]\centering
 	   \includegraphics[width=0.45\textwidth,keepaspectratio]{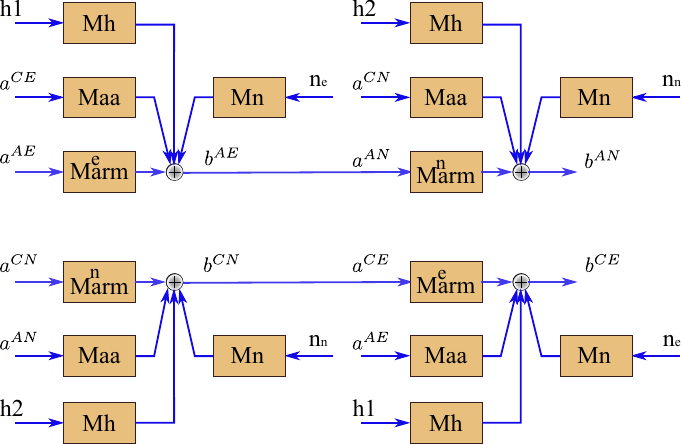}
    	   \caption{Block diagrams of a polarizing Sagnac
           interferometer. Each light field corresponds to that has shown in
           Fig.\,\ref{fig:pbs_config}\,(\rmnum{1}) and each block follows the
           definition in
           Appendix \ref{apps:BD}. The mechanical displacements
           $x_{1,2}$ due to the radiation pressure force of the light circulating in
           the cavities, occur only at the end mirrors. Any light
           transmitted by the end mirror (EM) is considered as an optical
           loss. For a lossless cavity, $R_e=1$, $T_e=0$. QWPs are assumed to
           be perfect for polarization rotation. Please note that ${\bf M}_{\rm
           arm}^{n}$ is exactly the same as ${\bf M}_{\rm arm}^{e}$; the
           different subscripts simply indicate which light field enters which
           arm first. }
   	   \label{fig:BD_sag}
	\end{figure}

For the implementation of the block diagram concept, the Sagnac interferometer
[see Fig.\,\ref{fig:pbs_config}\,(\rmnum{1})]
is split into several principal parts as shown by the blocks in
Fig.\,\ref{fig:BD_sag}. Each block is defined by its transfer
function (TF), where ${\bf M}_{\rm arm}$ is the TF of arm cavity, ${\bf M}_{\rm h}$ is the
GW signal
TF and ${\bf M}_{\rm n}$ is noise
TF which is induced by optical losses.
All the blocks used here are derived in Appendix \ref{apps:BD}.
Following the propagation path of the vacuum field through the Sagnac interferometer we can
represent the input-output relation. 
More specifically, we can derive the propagation relation between $g$
and $q$ [see Fig.\,\ref{fig:pbs_config}\,(\rmnum{1})] in terms of optical losses
(all the fields shown in the block diagrams have been denoted correspondingly in the schematic).
Following the
two signal
flows in Fig.\,\ref{fig:BD_sag}, we obtain
\begin{align}
\nonumber b^{CE}-b^{AN} = \;&{\bf M}_{\rm arm}(a^{CE}-a^{AN})+{\bf M}_{\rm aa}(a^{AE}-
a^{CN})\\
& +{\bf M}_{\rm h}(h_1-h_2)+{\bf M}_{\rm n}(n_e-n_n),\\
\nonumber a^{CE}-a^{AN}= \;&b^{CN}-b^{AE}\\
\nonumber =\;& {\bf M}_{\rm arm}(a^{CN}-a^{AE})-{\bf M}_{\rm aa}(a^{CE}-a^{AN})\\
\,&-{\bf M}_{\rm h}(h_1-h_2)-{\bf M}_{\rm n}(n_e-n_n),
\end{align}
where ${\bf M}_{\rm arm}$, ${\bf M}_{\rm h}$, and ${\bf M}_{\rm n}$ are
defined in Eqs.\,(\ref{equ:BDinoutlossycavity_ana}) and
(\ref{equ:BDinoutlossycavity_format}), and
\begin{equation}
{\bf M}_{\rm aa}=e^{2i\phi_{arm}}\left[\begin{array}{cc}0 & 0\\
-\kappa_{arm}&0\end{array}\right] 
\end{equation}
is induced by the two polarized fields circulating inside the arm cavities.
Combining the junction equations of light fields $g, q$, dimensionless
GW strain $h$ and additional optical losses induced quadrature $n$ at the
central BS
{\small
\begin{align}
&g=\frac{a^{AE}-a^{CN}}{\sqrt{2}},\qquad q=\frac{b^{CE}-b^{AN}}{\sqrt{2}}, \\
&h=h_1-h_2, \qquad\quad\;\;\, n=\frac{n_e-n_n}{\sqrt{2}},
\end{align}
}the input-output relation of a Sagnac interferometer with a perfect PBS and
optical losses in the arm cavities is
{\small
\begin{equation}
 q = {\bf M}_{\rm sag}g+{\bf H}_{\rm sag}h+{\bf N}_{\rm sag}n
	\label{equ:BDinoutlossysag_ana}, \end{equation}
	}where
{\small\begin{align}
\nonumber&{\bf M}_{\rm sag}=e^{2i\phi_{sag}}\left[ \begin{array}{cc}
1&  0\\
-\kappa_{sag} & 1  \end{array} \right],\qquad
{\bf H}_{\rm sag}=e^{i\phi_{sag}}
\frac{\sqrt{2\kappa_{sag}}}{h_{SQL}}\left[ \begin{array}{c}
0\\
1  \end{array} \right] \\
\nonumber&{\bf N}_{\rm sag}=e^{i\phi_{sag}}\sqrt{T_e}\sqrt{\frac{\kappa_{sag}}
{\kappa}}\left[ \begin{array}{cc}
1&  0\\
N& 1  \end{array} \right],\\
\label{equ:sagpara}&N=e^{2i\phi_{arm}}\kappa_{arm}-e^{i\phi_{arm}-i\Omega
\tau}\sqrt{\frac{\kappa \kappa_{arm}}{T_i}}\,.
\end{align}}%
The corresponding parameters, which follow the same definitions as in~\cite{Chen03}, are
{\small
\begin{align}
\nonumber&\kappa_{arm} = \frac{T_i\kappa}{1-2\sqrt{R_i}\cos(2\Omega\tau)
+R_i}, \qquad h_{SQL}=\sqrt{\frac{8\hbar}{m\Omega^2L^2}},\\
\nonumber&\phi_{arm} = \arctan{\left(\frac{1+\sqrt{R_i}}{1-\sqrt{R_i}}\tan{\Omega\tau}\right)},\quad \kappa=\frac{8I_c\omega_0}{mc^2\Omega^2},\quad \tau = \frac{L}
{c},\\
\label{equ:sagM}&\phi_{sag} = 2\phi_{arm}+\frac{\pi}{2},\qquad\kappa_{sag}
=4\kappa_{arm}\sin^2\phi_{arm},
\end{align}
}where $h_{SQL}$ is the \textit{standard quantum limit} (SQL), $R_{i}$ and
$T_{i}$ are the reflectivity and transmissivity of the arm cavity input
mirror, $I_c$ is the circulating power inside the arm cavities, $L$ is the
length of arm cavities, $\omega_0$ is the laser frequency, $\Omega$ is the GW
signal angular frequency, and $m$ is the reduced mass of the test masses. We have included
the full equations
for both lossless and lossy cavities in Appendix\,\ref{apps:BD}.
	
With the relation between $g$ and $q$ [Eq.\,\eqref{equ:BDinoutlossysag_ana}],
we can substitute the
Sagnac interferometer matrices and signal recycling (SR)
mirror parameters into Eq.\,(\ref{equ:BDinoutcavity}) to get the full
input-output relations ($g^{sr}$ and $q^{sr}$) of a perfect Sagnac
interferometer with SR in agreement with~\cite{Chen03}.
The ET xylophone design allows an
independent parameter optimization of the high frequency
sensitivity. The SR mirror, which helps to
improve and optimize
the quantum noise of a Sagnac interferometer at high frequencies at the cost of
low-frequency sensitivity, in the ET-LF case is no
longer necessary.
A setup without signal recycling will reduce the complexity
for the control systems significantly. Therefore in the rest of the
paper we discuss a setup without signal recycling as shown in
Fig.~\ref{fig:Sagnac_field_PBS}.
	\begin{figure}[!t]\centering
 	   \includegraphics[width=0.45\textwidth,keepaspectratio]{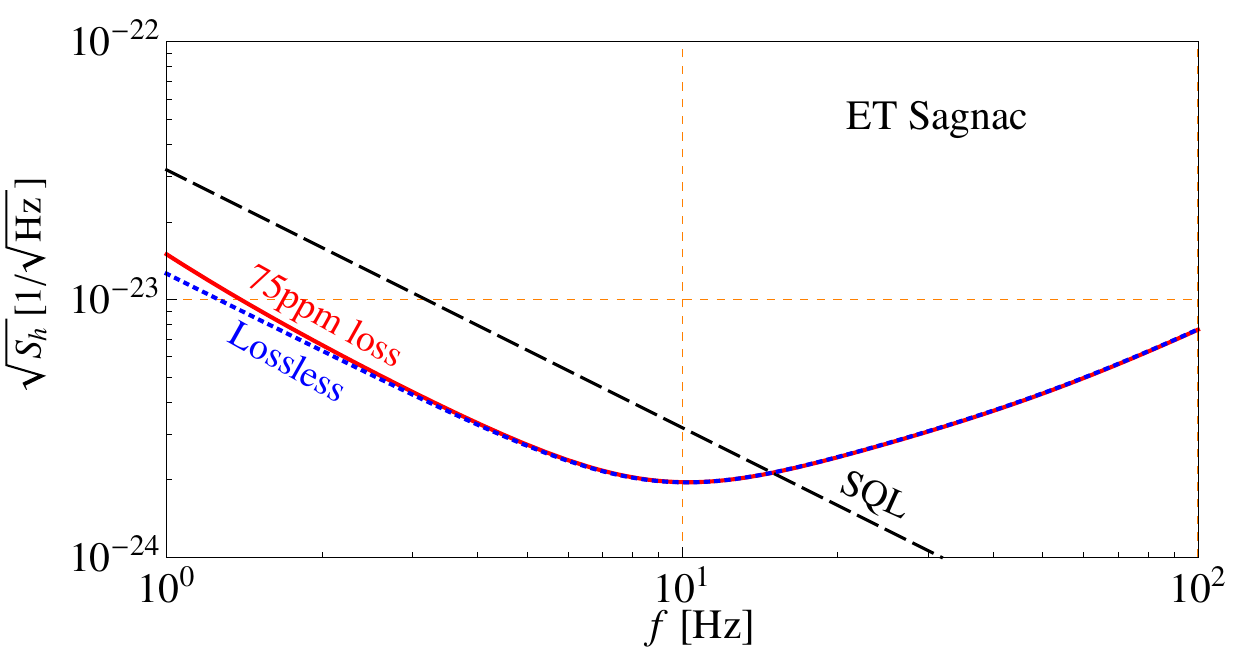}
    	   \caption{Plots showing the quantum NSD of a
	   polarizing Sagnac interferometer with the original ET-LF
	   Michelson parameters~\cite{ET11} (See Table\,\ref{tab:ET}):
	   10\,km long arm cavities,
	   10\,dB unique angle squeezing input and 18\,kW cavity
	   circulating power. The solid line is the polarizing Sagnac
	   interferometer with the same cavity round-trip loss being
	   considered, 75\,ppm. The dotted plot shows the quantum noise
	   spectrum of a perfect lossless polarizing Sagnac interferometer. The
	   dashed black line is the SQL defined by
	   Eq.\,(\ref{equ:sagM}). It will be shown as a reference in all the NSD plots
	   henceforth.}
   	   \label{fig:spe_sag_loss}
	\end{figure}

Given the above input-output relation and 
the NSD definition in Appendix\,\ref{app:NSD}, the quantum NSD of the 
Sagnac interferometer is
{\small
\begin{equation}\label{S_h}
S_h=\frac{e^{2r_p}(\cot\zeta-\kappa_{sag})^2+e^{-2r_p}}{2\kappa_{sag}}h^2_{SQL}+S_n,
\end{equation}}%
where the term $S_n$ comes from the optical losses and 
$r_p$ is the squeezing factor,
i.e., a 10\,dB phase squeezing corresponding to $r_{p}= 0.5\ln10$. By inserting the original ET-LF Michelson parameters~\cite{ET11} (summarized
as Michelson values in Table.\,\ref{tab:ET}), 
the sensitivity of a Sagnac interferometer with or without optical losses
taken into account can be obtained as shown in
Fig.\,\ref{fig:spe_sag_loss}. Here the
homodyne detection angle is chosen as
\begin{equation}\label{zeta_opt}
\zeta=\zeta_{opt}={\rm arccot} \left(\kappa_{sag}|_{\Omega\rightarrow 0}\right)
\end{equation}
to improve the low-frequency sensitivity by canceling the radiation pressure noise, which can be seen from Eq.\,\eqref{S_h} and the fact that $\kappa_{sag}$ is nearly a constant at low frequencies.

\begin{figure}[t]
     \centering
   	 \includegraphics[width=0.45\textwidth,keepaspectratio]{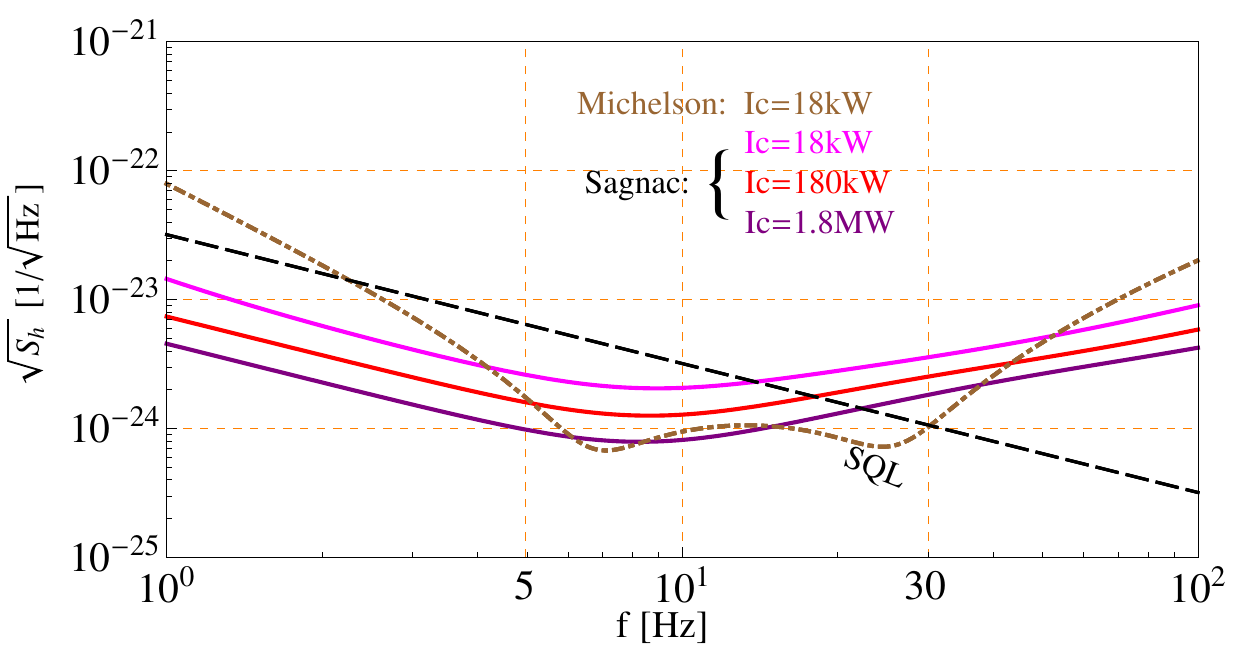}
    	   \caption{Plots showing the quantum NSD of a lossless
	   Sagnac interferometer with different powers circulating inside the
           arm cavities (${\rm I_c}$). The sensitivity peak is chosen around
           10\,Hz by adjusting the input power and the reflectivity of the cavity input
           mirror. The sensitivity curve for
           the ET-LF Michelson interferometer is
           shown as the dot-dashed line for comparison.}
   	   \label{fig:spe_sag_power}
	\end{figure}

ET-LF is not concerned with quantum noise reduction above 32\,Hz (covered
by ET-HF).
A Sagnac interferometer naturally provides good quantum-noise 
performance at low frequencies, and thus represents a good alternative 
for ET-LF. It should be noted that if the Sagnac configuration
were used with the same circulating power as the original ET-LF
Michelson, the new configuration would have a reduced peak sensitivity between 5 and
30\,Hz.
However, it can achieve a better quantum-noise limited sensitivity
below 5\,Hz as shown in
Fig.\,\ref{fig:spe_sag_power} (magenta curve),
providing the opportunity to
further improve the detectors performance by reducing other
limiting noise sources via upgrades or improvements of
subsystems of the detectors. Further technical study
is needed to trade off the worse peak sensitivity against the
inherent better low-frequency sensitivity and the much lower
complexity of the Sagnac configuration.
However, the peak sensitivity of the Sagnac could be easily
improved by increasing the laser power as indicated in
Fig.\,\ref{fig:spe_sag_power}.  We will
discuss this option briefly below.

From Fig.\,\ref{fig:pbs_config}\,(\rmnum{1}), we see that both polarizations contribute to the
radiation pressure force on the end mirrors (EM) and the circulating
power inside the arm cavity intrinsically determines the behavior of the
quantum noise.
A better low-frequency sensitivity can be achieved using
an increased circulating power $I_c$ inside the arm cavity. This is due to the fact
that the quantum noise level
is determined by the shot noise which is inversely proportional to the
laser power, while the radiation pressure noise at low frequencies is canceled 
out by choosing the optimized homodyne detection angle
$\zeta_{opt}$. We can achieve a better sensitivity by increasing the cavity circulating
 power as shown in Fig.\,\ref{fig:spe_sag_power}. Here, in order to optimize the 
 sensitivity peak around 10\,Hz, we also simultaneously tune the reflectivity of the 
 arm cavity input mirror, which modifies the detection bandwidth. 

However, any increase in circulation power will increase the mirror
and suspension thermal noise due to the finite mirror absorption, which 
in turn increases the temperature. A full optimization of such a design is
beyond the scope of this
paper, but we can gain some insight by discussing the scaling of the
thermal noise with temperature.
According to the simple analytical model given in~\cite{ET11_tn}, the
mirror temperature for the increase in light power by a factor of 10 would be doubled to
about 20\,K. For the increased mirror temperature,
we would expect an increase in  mirror thermal noise by a factor of
$\sqrt{2}$ and the same for the suspension thermal noise. This would
still remain below the quantum noise and thus not significantly
change the detector sensitivity.

\subsection{Sagnac interferometer with imperfect PBS }\label{sec:BadPBSsag}
\begin{figure}[t]\centering
 	   \includegraphics[width=0.4\textwidth,keepaspectratio]{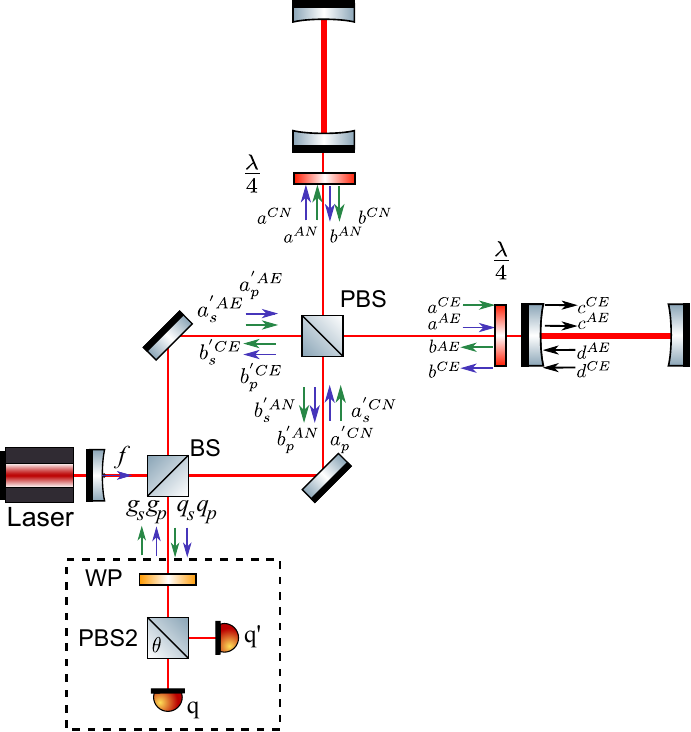}
    	   \caption{Schematic of a polarizing Sagnac interferometer with an
           imperfect PBS. The input beam is still purely
           {\it p}-polarized. The output has the coupled
           orthogonal {\it s}-polarized field due to the PBS
           imperfection. We also refer to this {\it s}-polarized field as the
           leakage from the PBS, which is the Michelson response. Homodyne
           detection is achieved as shown in the dashed box by using a wave plate
           (WP) and another PBS, which will be detailed in Sec.\,\ref{sec:SagMich}. 
           Squeezed vacuum is injected at the detection port. }
   	   \label{fig:Sagnac_field_PBS}
	\end{figure}

In order to propose a realistic configuration we have to consider the
finite extinction ratio of the PBS and understand the effects
of the light fields that are coupled into the other ports
due to that effect.
A good quality tabletop cubic
PBS has typical
$\eta_{p, s}=1/100\sim1/1000$ for {\it p}-polarized field
transmission
and {\it s}-polarized field reflection. Applied to a Sagnac interferometer, this
results in an output containing both orthogonally polarized fields, as shown
in Fig.\,\ref{fig:Sagnac_field_PBS}.

Following the same procedure discussed for the ideal case
(see Sec.\,\ref{sec:PerfectPBSsag}), we will investigate the
input-output relation with combined {\it s}-polarized and
{\it p}-polarized beams and {\it s}-polarized and {\it p}-polarized vacuum
fluctuations. In Appendix\,\ref{apps:PBS}, we derive the polarized
beam relations when both orthogonally polarized beams are
incident and outgoing at all ports of the imperfect PBS. Based
on those equations, we obtain the input-output relations of the
imperfect Sagnac interferometer.
We decided to perform the study of an imperfect optical layout for
an increased circulating power of $I_c=180$\,kW. This is best
suited to show the limits of a potential final implementation, while
the qualitative results are the same as for the low power scenario.
Apart from increasing the cavity circulating power
(summarized as Sagnac values in Table\,\ref{tab:ET}), we keep all of the
parameters proposed for the original ET-LF Michelson
interferometer.
	
\begin{figure}[t]\centering
 	   \includegraphics[width=0.45\textwidth,keepaspectratio]{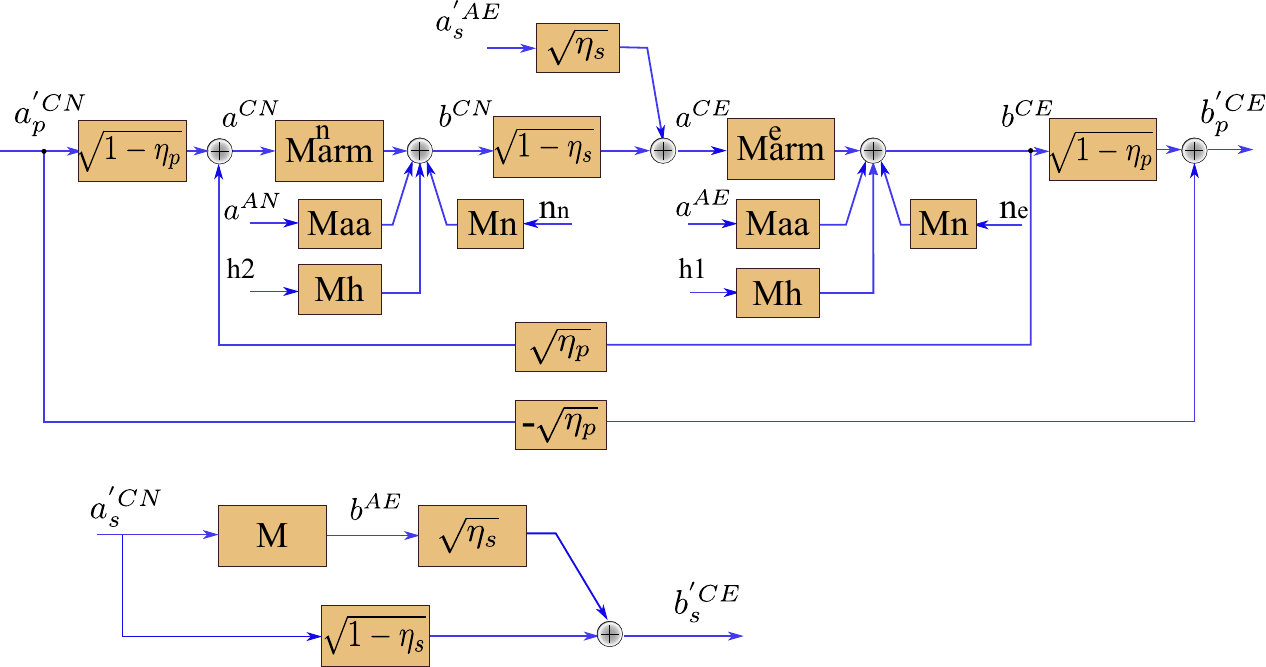}
    	   \caption{A block diagram of a polarizing Sagnac
           interferometer with finite extinction ratios $\eta_{p,s}$
           of the PBS. Block ${\bf M}$ is defined as $\sqrt{\eta_s}\sqrt{\eta_p}
           {\bf M}_{\rm CLG}{\bf M} _{\rm sag}$ and
           other the blocks keep the same as the perfect case in
           Fig.\ref{fig:BD_sag}. A
           closed loop is formed due to the effective reflection
           leakage of the {\it p}-polarized beam. The output
           contains both polarizations.}
   	   \label{fig:BD_sag_imperfect}
	\end{figure}

It has been recognized that the losses (i.e., absorption, scattering
loss) at the BSs, including central BS and PBS, have small effect on
the entire sensitivity
curve \cite{Chen03}. We hence ignore the losses at the BSs in our calculation, but
only focus on the different polarized beam leakages. A block
diagram similar to Fig.\,\ref{fig:BD_sag} can be obtained showing
a polarizing Sagnac interferometer using an imperfect PBS with extinction
ratios $\eta_s$ and $\eta_p$ (see
Fig.\,\ref{fig:BD_sag_imperfect}).
The input-output relations, shown in the block diagram in
Fig.\,\ref{fig:BD_sag_imperfect}, are highly symmetric with respect to
the order in which the vacuums enter the arms. Here, we illustrate the
input-output
relation of the light field firstly entering the \textit{vertical arm} (see
Fig.\,\ref{fig:Sagnac_field_PBS}). The
opposite field can be derived with an analysis similar to the perfect case in
Sec.\,\ref{sec:PerfectPBSsag}. With
conjunction equations (for both quadratures)
{\small
\begin{align}
g_p=\frac{  a_p^{'AE}-a_p^{'CN}}{\sqrt{2}}, \qquad g_s=\frac{  a_s^{'AE}-a_s^{'CN}}
{\sqrt{2}},\\
q_p=\frac{  b_p^{'CE}-b_p^{'AN}}{\sqrt{2}}, \qquad q_s=\frac{  b_s^{'CE}-  b_s^{'AN}}
{\sqrt{2}},
\end{align}
}we find (up to the order of $\sqrt{\eta_p}$ and $\sqrt{\eta_s}$),\footnote{Here, we  keep the leading order up to $\sqrt{\eta_{p,s}}$, so that the main features of the Sagnac and Michelson response of both polarizations maintain. Two complete equations are shown in Appendix \ref{app:full}.}
{\small\begin{align}
\nonumber  q_p =
	&(-\sqrt{\eta_p}{\bf I}+
	{\bf M}_{\rm CLG}{\bf M}_{\rm sag})
	 {g}_{p}+\sqrt{\eta_s}{\bf M}_{\rm CLG}{\bf M}_{\rm arm} {g}_{s}\\
\label{equ:inout_p}&+{\bf M}_{\rm CLG}{\bf H}_{\rm sag}h
         +{\bf M}_{\rm CLG}{\bf N}_{\rm sag}  n\,,\\
\nonumber {q}_{s} =
	&- {g}_{s}
         +\sqrt{\eta_s}{\bf M}_{\rm CLG}{\bf M}_{\rm arm}
	 {g}_{p}\\
\label{equ:inout_s}&+\sqrt{\eta_s}{\bf M}_{\rm CLG}{\bf H}_{\rm arm}h
         +\sqrt{\eta_s}{\bf M}_{\rm CLG}{\bf N}_{\rm arm}
    {n}\,,
\end{align}
}where the corresponding matrices and parameters are defined in Eqs.\,(\ref{equ:sagpara}) and (\ref{equ:sagM}), and the closed loop gain due to reflection of the PBS with finite $\eta_{s,p}$ is given by
{\small
\begin{equation}
{\bf M}_{\rm CLG}=\left[{\bf I}-\sqrt{\eta_p(1-\eta_s)}{\bf M}_{\rm sag}\right]^{-1},
\end{equation}
}of which the influence on the overall response of the interferometer is minor---${\bf M}_{\rm CLG}\approx {\bf I}$,
as $\eta_{p,s}\ll 1$.

Given the parameters we shall use,  the above input-output relation can be well approximated as (with negligible error)
{\small\begin{align}
  {q}_{p}&\approx{\bf M}_{\rm sag} {g}_{p}+\sqrt{\eta_s}  {\bf M}_{\rm arm}{g}_{s} + {\bf H}_{\rm sag}h\,, \\
 {q}_{s}&\approx -{g}_{s}+\sqrt{\eta_s}{\bf M}_{\rm arm}g_p+\sqrt{\eta_s}{\bf H}_{\rm arm}h\,.
\end{align}
}These relations reveal two interesting facts arising from the finite extinction ratio
of the PBS: (i)
the vacuum fluctuations for both polarized fields are mixed. In particular, for the
output of the {\it p}-polarized field, the {\it s}-polarized vacuum $g_s$ induces a radiation
pressure noise which has the same frequency dependence as the one in a
typical
Michelson interferometer (see ${\bf M}_{\rm arm} g_s$ term). As we shall see, this
will degrade the low-frequency sensitivity of the Sagnac interferometer; (ii) the
{\it s}-polarized output gains a Michelson-type response (see the
${\bf H}_{\rm arm}h$
term). Even though such a response of  $q_s$ to the GW signal is
negligible, as $\sqrt{\eta_s}\ll 1$,  we can utilize it to create a
local oscillator field
by inducing a small offset $\Delta L$ of the two arms, namely
{\small\begin{equation}q_s\approx -g_s+\sqrt{\eta_s}{\bf M}_{\rm arm}g_p+ q_{\rm LO}\end{equation}}with
{\small\begin{equation}
q_{\rm LO}=\sqrt{\eta_s}{\bf H}_{\rm arm}\Delta L/L\,.
\label{equ:dc}\end{equation}
}%
This produces a LO in a way similar to the DC readout scheme that will be
implemented in advanced GW detectors, see Sec.\,\ref{sec:DC_readout}.

To mix this LO with $q_p$ for the homodyne detection, another PBS at the output
port with adjustable optical axis  is necessary, as these two outputs $q_s$ and
$q_p$ have orthogonal polarizations. The corresponding scheme is shown in the
dashed box in Fig.\,\ref{fig:Sagnac_field_PBS}. By adjusting the optical axis of
the
PBS, we can tune the \textit{detection ratio angle} $\theta$.  The resulting two
outputs after such a PBS are given by
{\small\begin{align}
\label{equ:q_theta}q^{\theta}&= q_p\cos \theta+q_s
\sin \theta\,,\\
\label{equ:s_theta}q'^{\theta}&= q_p\sin \theta+q_s
\cos \theta\,.
\end{align}
}For a small $\theta$, the majority response of $q^{\theta}$ is still the Sagnac
signal with $q_s\sin\theta$ providing the LO.
The detailed detection scheme as well as the usage of
$q'^{\theta}$ for optics position control will be outlined in Sec.\,\ref{sec:DC_readout}.

By tuning the phase of the LO, we can measure the $\zeta$ quadrature\footnote{The homodyne detection angle $\zeta$ is determined by
the
relative phase difference between $q_p$ and $q_s$, which can be controlled
by defining the thickness of a wave plate before the PBS as shown in the dashed box
of Fig.\,\ref{fig:Sagnac_field_PBS}.} of
$q^{\theta}$, and the final output is given by (for small
$\theta$)
{\small\begin{align}\nonumber
q &= q^{\theta}_{1} \cos \zeta+q^{\theta}_2\sin\zeta\,\\
&\approx  (q_{p_1} + \theta\, q_{s_1})\cos\zeta +(q_{p_2} + \theta \,q_{s_2})
\sin\zeta\,.
\end{align}}%
We can obtain the strain $h$-referred NSD of
$q$ as
{\small\begin{align}\nonumber
S_h=& \frac{e^{2r_p}(\cot\zeta -\kappa_{sag})^2 +e^{-2r_p} }{2\kappa_{sag}} h^2_{SQL}\\
&+ \frac{e^{2r_s}[(\sqrt{\eta_s}+\theta)\cot\zeta -\sqrt{\eta_s}\kappa_{arm}]^2
+e^{-2r_s}\theta^2}{2\kappa_{sag}}h^2_{SQL}\,, \label{equ:spectrum}
\end{align}
}where $r_p$ and $r_s$ are the squeezing factors and we have assumed that different frequency-independent squeezed light
can be injected for both polarizations. Notice that
(i) the term in the first line of Eq.\,(\ref{equ:spectrum}) corresponds to the usual NSD for a
Sagnac interferometer, and $\kappa_{sag}$ is nearly flat at frequencies lower
than the arm cavity bandwidth. As mentioned earlier [see Eq.\,\eqref{zeta_opt}], 
by choosing the correct
homodyne detection angle $\zeta$, we can remove the low-frequency radiation
pressure noise; (ii) the term in the second line arises from the finite extinction
ratio
$\eta_s$ and the detection ratio angle $\theta$. As mentioned earlier,
this increases the low-frequency radiation pressure noise due to the frequency
dependence of $\kappa_{arm}\propto \Omega^{-2}$ (higher at lower
frequencies)
in contrast to $\kappa_{sag}\propto \Omega^0$ at low frequencies. To mitigate its influence, one
apparent approach is to minimize $\eta_s$ and $\theta$. An alternative
method is to inject amplitude squeezed light for the {\it s}-polarization, namely $r_s<0$.

		\begin{figure}[b]\centering
 	   \includegraphics[width=0.4\textwidth,keepaspectratio]{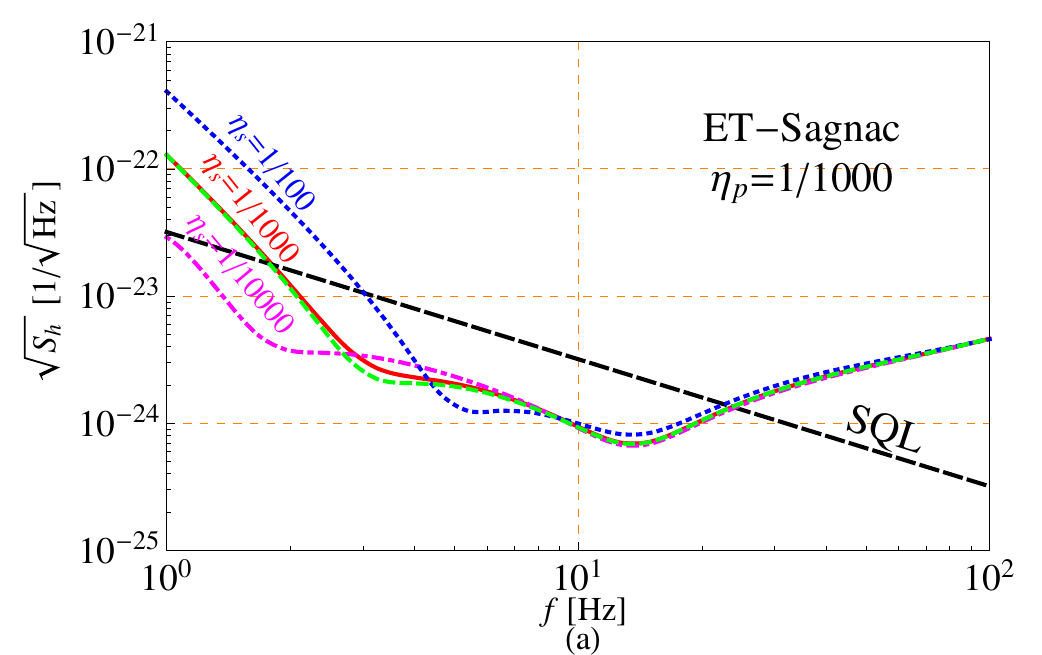}
	   	 \includegraphics[width=0.4\textwidth]{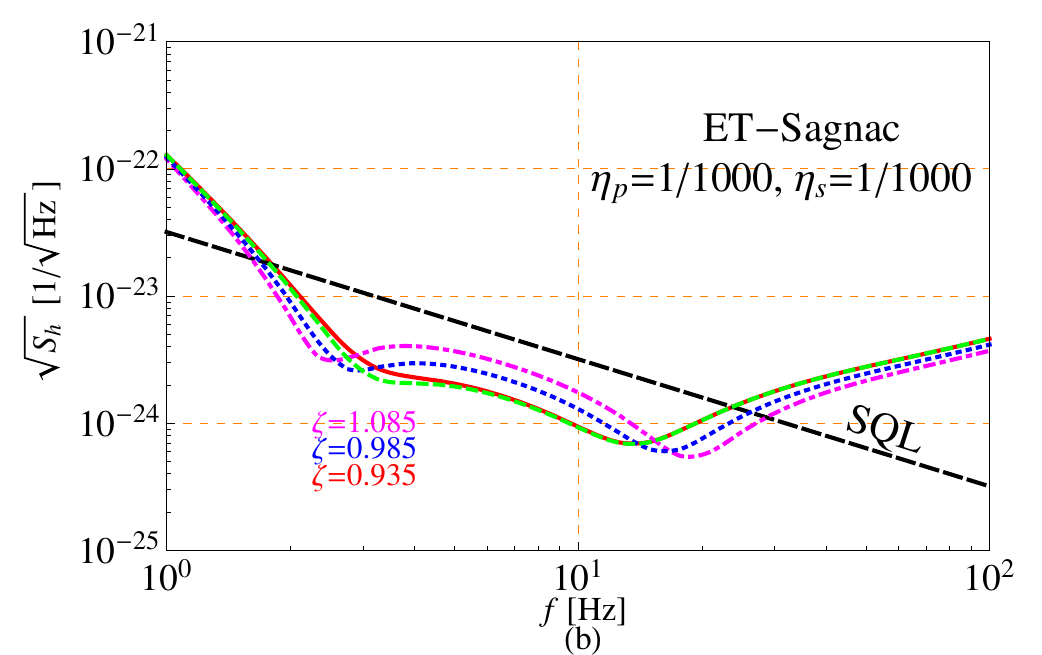}
		 \includegraphics[width=0.4\textwidth]{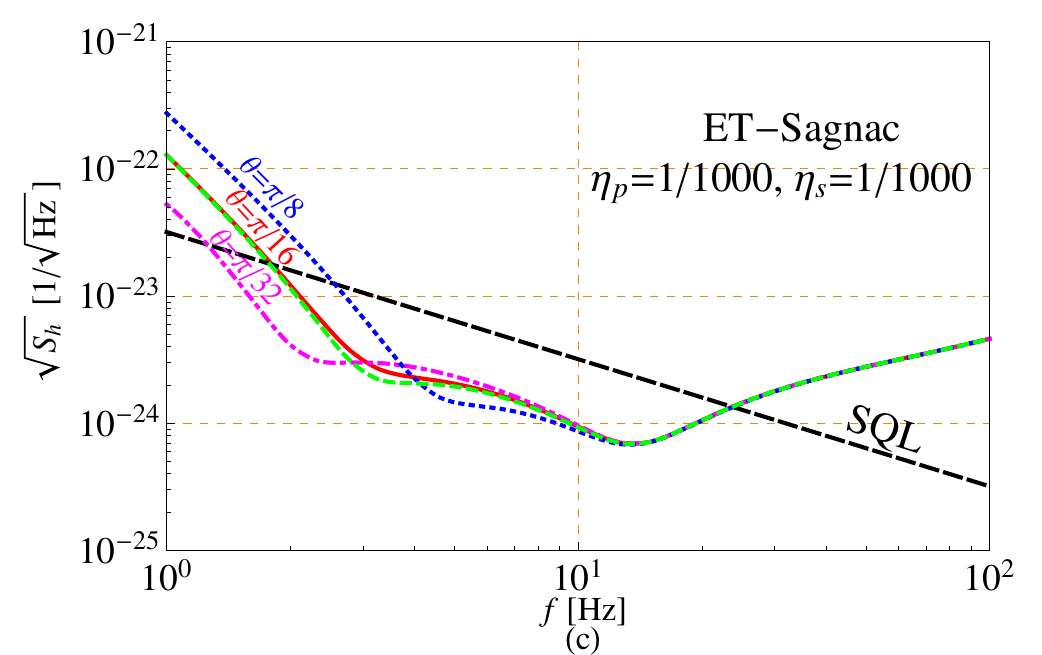}
    	   \caption{Plots showing the quantum NSD of polarizing Sagnac
           interferometers with different specifications for selected parameters.
           The configuration assumes the Sagnac
           interferometer parameters shown in
           Table~\ref{tab:ET}. All plots are based on a default
           parameter set with detection angle $\theta=\pi/16$,
           homodyne detection angle $\zeta=0.935$ and
           $\eta_s=\eta_p=1/1000$. Part (a)
           shows the impacts of different extinction ratios
           $\eta_s$. $\eta_p$ is fixed, as it has little
           influence to the final results. Part (b)
           illustrates a narrow band quantum noise mitigation via a
           homodyne detection angle selection. Part (c) shows the noise
           spectrum for
           different detection ratio angles. The dashed green curves in all plots 
           which are almost overlapped
           with those red curves (lossless Sagnac) 
           illustrate the sensitivity of a lossy Sagnac interferometer when 75\,ppm 
           arm cavity round-trip loss being considered.}
	      	   \label{fig:spe_sag_bad}
	\end{figure}

   \begin{table}[t]
    \begin{center}
    \begin{tabular}{lcc}
    \hline
    \hline
    Parameter& Michelson & Sagnac \\
    \hline
    Arm length (L)  & 10\,km & 10\,km   \\
    Input Power (after IMC) & 3\,W  & 15\,W  \\
    Input Power at BS & 138\,W & 690\,W\\
    Arm Cavity Power ($I_c$)& 18\,kW & 180\,kW   \\
    Temperature & 10\,K & 20\,K   \\
    Mirror Mass & 211\,kg & 211\,kg   \\
    Laser Wavelength  & 1550\,nm & 1550\,nm   \\
    SR Detuning Phase & 0.6  & ...  \\
    SR Transmittance & 20\% & ...   \\
    Filter Cavities & 2 $\times$ 10\,km & ...  \\
    Squeezed Level & 10\,dB & $\pm$10\,dB \\
    Scatter loss per surface & 37.5\,ppm & 37.5\,ppm\\
    \hline
    \end{tabular}
    \end{center}
    \caption{A table summarizing the parameters of ET-LF
      interferometers. The Michelson values are taken from the
      design study \cite{ET11}. The Sagnac parameters proposed here
      refer to the high-power scenario with a circulating power of
      $I_c=180$\,kW which gives the sensitivity shown in Fig.\,\ref{fig:spe_sag_power}.}
    \label{tab:ET}
    \end{table}

In Fig.\,\ref{fig:spe_sag_bad}, we show the resulting quantum NSD for different
specifications of the parameters.
We have assumed a default parameter base: (1) a small
detection ratio angle $\theta=\pi/16$, (2) a homodyne detection angle
$\zeta=0.935$, which gives an optimal sensitivity at a frequency around 10\,Hz,
(3) reasonable PBS extinction
ratios $\eta_s=\eta_p=1/1000$, and (4) phase squeezing for the
{\it p}-polarization, $r_p = 0.5 \ln10$
and amplitude squeezing for the {\it s}-polarization, $r_s =-0.5 \ln 10$.
Figure~\ref{fig:spe_sag_bad} (a) shows the effects when various
quality
PBSs are implemented with $\theta = \pi/16$ and
$\zeta=0.935$. In Fig.\,\ref{fig:spe_sag_bad} (b) we keep
$\eta_s=\eta_p=1/1000$ and $\theta = \pi/16$, and change the homodyne
detection angle $\zeta$. Figure\,\ref{fig:spe_sag_bad} (c) gives
a possible detection ratio optimization when the leakage influences
the quantum noise behavior. We have found that (\Rmnum{1}) the low
frequency quantum noise sensitivity greatly depends on the quality of
the {\it s}-polarized reflection extinction ratio $\eta_s$, but is
rarely impacted by the {\it p}-polarized transmission ratio $\eta_p$,
given that $\eta_{s,p}$ are both smaller than 1\%; (\Rmnum{2}) the
homodyne detection angle can be optimized to slightly mitigate a
narrow band quantum noise; (\Rmnum{3}) the losses have negligible
impact on the sensitivity as one can imagine that the influence of the
low-frequency losses are covered by the Michelson interferometer
response. This is confirmed in each curve in
Fig.\,\ref{fig:spe_sag_bad}, where the dashed green curve (illustrating the
lossy Sagnac sensitivity) is almost identical to the solid lossless red curve;
(\Rmnum{4}) the quantum noise behavior can be improved, as long as the
Michelson output ensures the minimum DC requirement of the photodiode,
which will be detailed in Sec.\,\ref{sec:DC_readout}.

\section{DC readout}\label{sec:DC_readout}
The optical readout of a GW signal at the output of
an interferometer requires a so-called local oscillator (LO), a
reference light field which beats with the signal field on the photodiode. Current detectors use a concept called
DC readout~\cite{Hild09b} in which
a part of the main circulating light field is directed into the
dark port for this purpose. This concept has the advantage
that the LO light is already prefiltered by the large baseline
interferometer, on-axis and phase locked to the signal field.

In a Michelson interferometer the amount of carrier leakage into the dark port can
be controlled with a small offset of the differential arm lengths.
In an ideal Sagnac interferometer however, the dark fringe is independent
of the mirror position. In the next section we investigate the possibility
of generating a DC readout LO for the polarizing Sagnac
interferometer. Additionally, we introduce a new method to select and control
the homodyne detection angle for such an interferometer.

\subsection{Required light level and homodyne detection angle}
The detected signal and the shot noise both scale as the square root
of the LO power. In order to reach shot noise
limited performance the light power in the LO
must be large enough to (\rmnum{1}) have the photodiode dark
noise lower than the shot noise and 
(\rmnum{2}) dominate over waste light from higher order modes and stray light on
the photodiode; it must also (\rmnum{3})
have the correct homodyne phase. 

The exact light power required in the LO thus depends on the technical
details of the interferometer implementation. The ET design
has not yet reached such a level of detail and instead uses
specifications for Advanced LIGO interferometers~\cite{adligo_design,
T1000298-v2} as guidelines. In the following section, we focus with our
investigation on the coupling of the fundamental mode into the detection port due
to an intentional dark fringe offset and an arm imbalance~\cite{Abbott05}. 

It is reasonable to assume that stray light and photodiode dark noise
will be similar in a Michelson and Sagnac interferometer. However, the coupling
of light into the detection port has different origins in the two cases. In a
Michelson interferometer it is caused by unequal losses in the two
arms, but in a Sagnac
interferometer it comes from a non-50:50 central BS. A 0.1\% deviation
from a perfect 50:50, which
we consider reasonable to expect from a technical point of view, would lead
coupling of $2\times 10^{-6}$ (relative to the circulating
power at the central BS). This is larger than the Advanced LIGO
estimate of $2\times 10^{-7}$, but as the Sagnac's optimum homodyne
angle ($\approx 53^\circ$) is further from $90^\circ$ (the signal) than the
Michelson one ($\approx 100^\circ$), it can tolerate a higher ratio of light field
(homodyne angle $0^\circ$ in both cases) to
LO~\cite{Abbott05}.  In the following discussion, we therefore assume the
same ratio of DC output power to central BS circulating power (which
we call $\gamma$) as in Advanced LIGO, $1.75\times 10^{-5}$.

Below we consider two
methods to achieve a
similar LO light power in the Sagnac interferometer: (\rmnum{1}) nonzero
area  Sagnac interferometers and (\rmnum{2}) PBS leakage.

\subsection{Sagnac area effect}\label{sec:SagnacArea}
A Sagnac interferometer with a nonzero area $\bf A$ responds to
rotation $\Omega$ as
{\small
\begin{equation}
\Delta L=\frac{4 {\bf A}\cdot {\rm \Omega} }{c},
\end{equation}
}where $c$ is the speed of light and $\Delta L$ is the same as shown in
Eq.\,(\ref{equ:dc}).
The ratio $\gamma$ is
{\small
\begin{equation}
\gamma=\sin^2 \left(\frac{2\pi\Delta L}{\lambda}\right).
\end{equation}
}When considering the Earth's rotation, this corresponds to
{\small
\begin{equation}
\gamma=1.75\times 10^{-5}\left(\frac{A}{1350\,\rm m^2}\right)^2\left(\frac{1550\,\rm nm}{\lambda}
\right)^2\left(\frac{\sin(\text{latitude})}{\sin(52^\circ)}\right)^2.
\end{equation}}%
The 1350\,$\rm m^2$ area required for $\gamma=1.75\times 10^{-5}$ would have to be folded (requiring extra mirrors)
to fit in the ET cavern; a 600\,$\rm m^2$ loop (about the largest size that
would fit as a four-mirror configuration) would have power fraction $\gamma=3.5\times
10^{-6}$.  This method is also likely to give a suboptimal homodyne angle, as the strength of the LO is fixed by the loop size, and the homodyne angle is set by the ratio of this fixed LO strength to the (typically unknown prior to construction)
light power due to an arm imbalance~\cite{Abbott05}.  Hence, we consider the PBS leakage method described in the next section to be more practical.

\subsection{PBS leakage light}\label{sec:SagMich}
As discussed in Sec.\,\ref{sec:BadPBSsag}, the leakage of an
imperfect PBS creates a Michelson response signal. Here
we consider the use of this field as the LO.
\begin{figure}[b]\centering
 	   \includegraphics[width=0.45\textwidth,keepaspectratio]{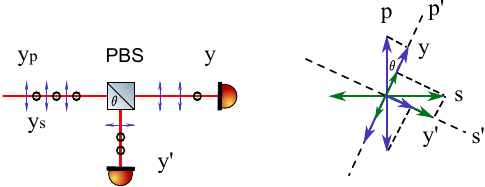}
    	   \caption{Diagrams illustrating a scheme of the outputs of two orthogonal polarized beams
	   with an optical axis rotated PBS. The input beams contain
           both polarizations: {\it p}-polarized beam (arrow) and {\it s}-polarized beam
           (circle). The left diagram shows the optical layout at the
           detection port; the right diagram illustrates the output details by rotating
           the PBS's polarization axes (p-s) by an angle $\theta$
           (p$'$-s$'$).}
   	   \label{fig:homodyne_off}
\end{figure}

If we measure two diagonal polarized beams after a
PBS which has rotated polarization axes as shown in Fig.\,\ref{fig:homodyne_off}, we can choose the strength of the
individual polarizations. With a rotation angle $\theta$, the output fields are
{\small
\begin{align}
y&=y_p\cos\theta+y_s\sin\theta,\\
y'&=y_p\sin\theta+y_s\cos\theta.
\end{align}
}We know that the Michelson signal from the polarizing Sagnac
interferometer is in the opposite polarization to the Sagnac
signal. The parasitic Michelson interferometer has a circulating power $\eta_s$
times the Sagnac's, and by setting it to bright fringe, we can send all of this to the
output, which can be used as a LO [see Eq.\,(\ref{equ:q_theta})]. The output ratio
from the Michelson leakage is
{\small
\begin{align}
\gamma=\eta_s\sin^2\theta=1.75\times 10^{-5}\left(\frac{\eta_s}{0.001}\right)\left(\frac{\sin\theta}
{0.13}\right)^2\,.
\end{align}
}From Fig.\ref{fig:spe_sag_bad} (c) we have found that a small detection ratio is preferred, 
we choose $\theta=\pi/24$ such that $\gamma=1.75\times 10^{-5}$. 

When the Michelson signal at the output is used as a LO for
detection, the homodyne detection angle $\zeta$ is determined by the relative
phase difference between the Sagnac signal ({\it p}-polarization) and the Michelson
signal ({\it s}-polarization). In principle, the two responses are naturally in-phase
at the detection port (see $q_s$ and $q_p$ in Fig.\,\ref{fig:spe_sag_bad}) and
a wave plate (WP) can be used to shift the phase
between them.
The required homodyne detection angle then can be defined by a
carefully chosen WP orientation and thickness. Compared to using an unknown
amount of light due to an arm imbalance to define the homodyne detection angle,
this method has obvious advantages. 
\subsection{Potential control of a Sagnac with PBS leakage}

In the current Michelson based GW detectors the positions of the
mirrors have to be carefully controlled via error signals generated for
different degrees of freedom (DoFs).  In the Michelson we control PRCL (the length of
the power recycling cavity); CARM (the common motion of the two arm cavities); DARM (the
differential motion of the arm cavities); MICH (the Michelson differential length) and SRCL (the
length of the signal recycling cavity).  The DARM error signal is
generated via DC readout whereas other
error signals are obtained by adding RF sidebands
to the input beam and demodulating these signals at various pickoff points within the detector.
Common pickoff points include the reflection from the detector (REFL),  between the
BS and ITM of one of the arms (POX/Y) or at the antisymmetric output of the interferometer (AS).

The following is a preliminary investigation into error signals for
controlling the proposed Sagnac interferometer.
First, we consider which DoFs we require control of in our Sagnac detector.
In the absence of a signal recycling mirror we do not require a
control signal for SRCL.  An initial consideration of the Sagnac output might suggest
less tightly required control of some of the DoFs, notably DARM and MICH, as the
light split at the BS travels through both arms, effectively keeping the interferometer
consistently on the dark fringe.  However, this will not be the case for a Sagnac acting as
a GW detector.  In this case we require tight control of the arms cavities to keep
them on resonance and for our realistic approach we require control of the parasitic
Michelson to prevent unreasonable fluctuations in our local oscillator.  Therefore we expect
to require good control signals for PRCL, CARM, DARM, and MICH.

\begin{figure}[t]\centering
 	   \includegraphics[width=0.39\textwidth]{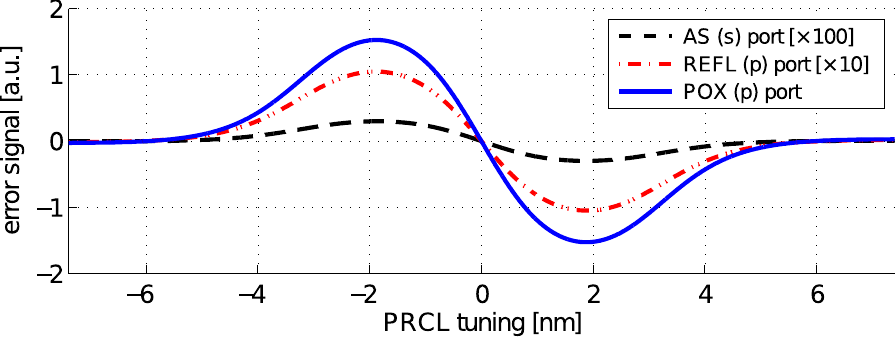}
	   	\includegraphics[width=0.39\textwidth]{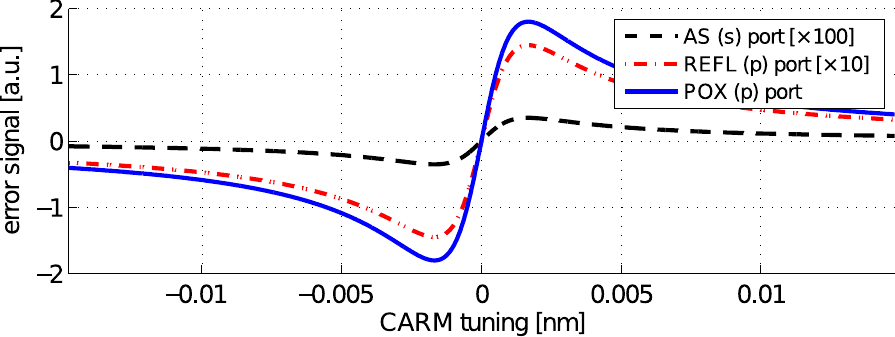}
		\includegraphics[width=0.39\textwidth]{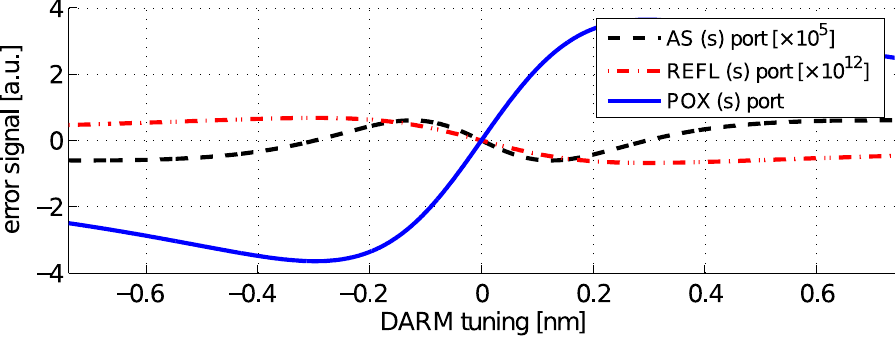}
		\includegraphics[width=0.4\textwidth]{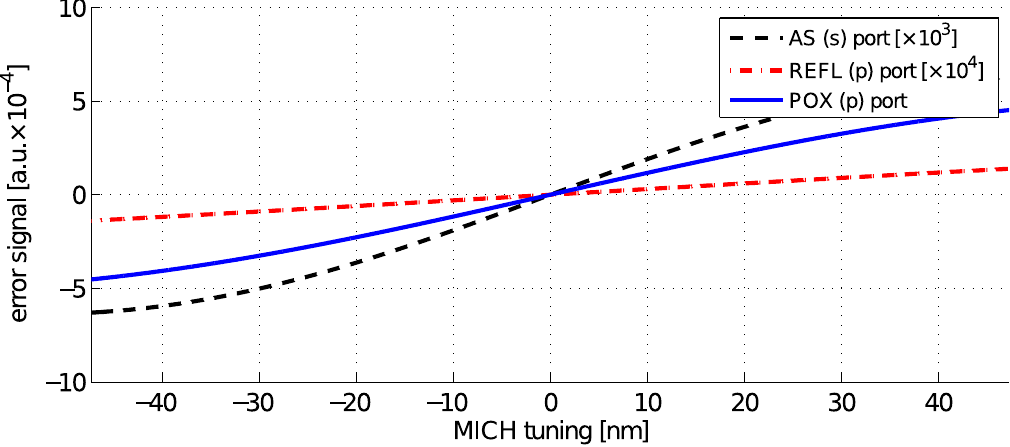}	
    	   \caption{Plots showing preliminary error signals generated in
	   \textsc{Finesse} for a realistic Sagnac detector, using ET-LF parameters
	   and $\theta=\pi/24$, $\zeta=0.935$, $\eta_s=\eta_p=1/1000$ and
	   with an increased arm cavity power of 180\,kW.
	   The different plots represent the signals for different degrees of
	   freedom (DoFs), from top to bottom: PRCL, CARM, DARM, and MICH.
	   For each DoF possible error signals are generated by applying
	   control sidebands at 1.25\,MHz and demodulating the signals at
	   three different ports: AS (antisymmetric), REFL (reflected), and
	   POX ($x$ arm pick off).  The polarization of each signal ({\it s} or {\it p})
	   is indicated in the plot labels.}
	      	   \label{fig:err_sigs}
	\end{figure}

The output field of our realistic Sagnac interferometer ($q$) is in
the {\it p}-polarization and contains mostly
the Sagnac response, including any GW signal, with a small portion of
light from the parasitic Michelson acting as a LO.
The combination of the Sagnac and Michelson responses is achieved via a PBS
and results in a second output signal, $q'$ [see Fig.\,\ref{fig:Sagnac_field_PBS}].
Unlike $q$, $q'$ is {\it s}-polarized and dominated by the Michelson signal,
containing just a small fraction of the Sagnac response.
This signal is not used as the readout of the detector but could be used to control
certain DoFs.  Using the interferometer simulation tool \textsc{Finesse}, the polarized
Sagnac was modeled using the realistic parameters $\theta=\pi/24$, $\zeta=0.935$ and
$\eta_s=\eta_p=1/1000$ with an increased circulating arm power of 180\,kW.
RF sidebands of 1.25\,MHz are added to the input beam.  Error
signals for the 4 degrees of freedom were produced by demodulating
the two polarization fields at the
AS,\footnote{In this case the AS port refers to the anti-symmetric port
of the Sagnac detector.  At this port the detector output ($q$) is
{\it p}-polarized and contains mostly the Sagnac signal.  The
{\it s}-polarized light contains mostly the Michelson signal and
therefore the AS port refers to the symmetric port for the Michelson
as we operate on the Michelson bright fringe.}
REFL, and POX ($x$ arm pickoff) ports.

In Fig.\,\ref{fig:err_sigs} we show some preliminary simulated error signals for the four DoFs;
PRCL, CARM, DARM and MICH; required for our realistic Sagnac.
The three ports investigated here demonstrate the potential for control of
such a detector, with several possibilities for each DoF.  This suggests a
control scheme for a realistic Sagnac with PBS leakage could be realized
in a similar manner to a conventional Michelson detector.


\section{Conclusion}\label{sec:conclusion}

It has previously been shown that a Sagnac interferometer without
filter cavities can
achieve a similarly low level of quantum noise at low frequencies as a
Michelson with filter cavities~\cite{Danilishin12}. We have built on
this premise, presenting an alternative topology for the Einstein
Telescope, replacing the Michelson interferometers of the low
frequency detectors with Sagnac interferometers.  Our scheme
employs polarizing optics (a PBS and QWP) to direct the beam whilst being
compatible with the current ET
infrastructure and avoiding the technical problems caused by long
ring-shape arm cavities.

We initially considered the performance of a Sagnac for ET-LF in the
case of a perfect PBS and investigated the effects of losses on the
performance, in terms of the quantum noise.  Using the ET-LF
parameters we found that the quantum-noise limited sensitivity
of the Sagnac is better below 5\,Hz but is reduced around the
peak between 5 and 30\,Hz. This reduction must be seen in the
context of a greatly reduced complexity of the system that
does not require filter cavities and signal recycling mirror.
We showed that by increasing
the circulating power by a factor of 10 we are able to achieve a comparable
quantum noise with a Michelson with filter cavities, between 5 and
30\,Hz, with even greater sensitivity below 5\,Hz and discussed the
feasibility of the higher power.  We also find that
including expected ET round-trip losses changes the quantum noise
curve very little in the frequency band we are concerned with.

 \begin{figure}[t]\centering
 	   \includegraphics[width=0.45\textwidth,keepaspectratio]{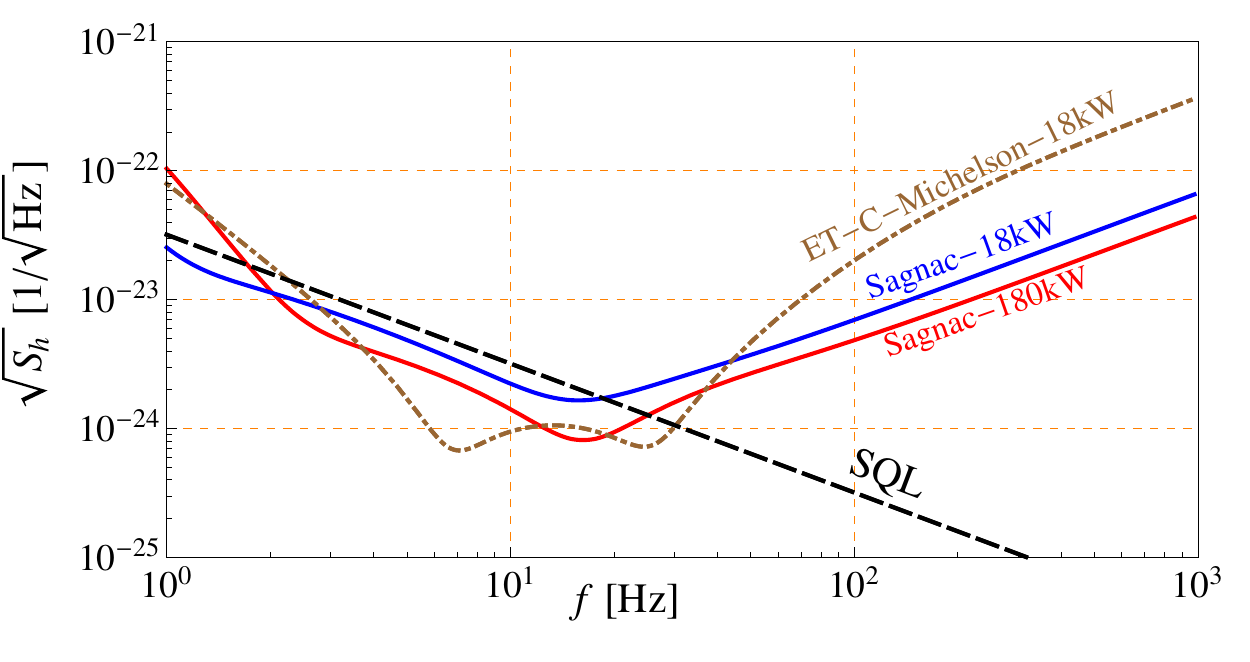}
    	   \caption{Plots comparing the quantum NSD of
           the proposed Sagnac topology (blue curve) against the ET-C
           Michelson
           topology based on same ET-LF Michelson parameters (See
           Table.\,\ref{tab:ET}).
           The Sagnac topology uses
           parameters
           $\theta=\pi/24$,
           $\zeta=0.935$ and $\eta_s=\eta_p=1/1000$ to ensure a
           LO for DC readout.
           Both Sagnac curves include the effects of optical losses, and
           the ET-C curve shown here does not
	  consider optical
           losses in the filter cavities.
           The red curve shows a Sagnac with a higher power, 180\,kW,
           circulating in the arm cavities, which
           would be a possible implementation to increase the peak sensitivity.}
	      	   \label{fig:spe_compare}
\end{figure}

The main part of our investigation involved considering the effects
and the possible operational advantages of a Sagnac with an imperfect
PBS, specifically with a finite extinction ratio.  By adapting our
model to include the effects of a realistic PBS we demonstrated that
the effect of a finite extinction ratio is described by the coupling
of a Michelson signal onto the Sagnac input-output relation, where the
signals of the Michelson and Sagnac are in the opposite
polarizations. We found that the quantum
noise of such an interferometer, at low frequencies, depends greatly
on $\eta_{s}$, the extinction ratio of the reflection of the
{\it s}-polarized beam and very little on $\eta_{p}$, the extinction ratio
of the transmission of the {\it p}-polarized beam. This is due to the fact that
{\it s}-polarized
vacuum fluctuations are directly coupled in. Amplitude squeezed {\it s}-polarized
squeezing light was revealed to be 
an effective approach to mitigate low-frequency
quantum noise. As the quantum behavior of the Michelson at low
frequencies is worse than the Sagnac, this combination of the two
signals results in a degraded sensitivity.  Again the impact of
losses on the quantum noise performance was found to be negligible.
We also found that the homodyne detection angle can be optimized to
provide a broadband quantum noise reduction.
Finally, we
presented the quantum noise curve for a realistic candidate for a
Sagnac interferometer for ET-LF (using an increased circulating power
but otherwise retaining the original ET parameters).
We consider a
case with extinction ratios $\eta_{s}=\eta_{p}=1/1000$ and achieve a
comparable quantum noise curve to a Michelson with filter cavities.
The implementation of such an interferometer in the Einstein Telescope
has significant implications for lowering costs and reducing the
complexity which occurs with additional filter cavities, as well as the SR mirror.
This topology also leaves room for further improvements and additional
technologies, i.e., better suspensions and cryogenic mirrors.

The quantum noise behavior of the proposed Sagnac interferometer can
be further improved by reducing the Michelson signal present in the
detector output, by means of reducing the detection ratio angle.
However, some of the Michelson signal is required to provide a LO
for homodyne detection. Given the ET parameters we proposed, the absolute value of the LO power always ensures a lower photodiode dark noise compared to the shot noise. For our purpose, we used the Advanced LIGO ratio of
DC output power to central BS circulating power,
$1.75\times10^{-5}$. We also considered using a nonzero Sagnac area
interferometer to provide the required bias, but have found this to be
impractical due to the large area required and suboptimal for homodyne angle
selection. The leaked DC light
provided by the nonperfect PBS is convenient for providing the LO and choosing homodyne detection angle. The required LO level is
achievable with realistic extinction ratios. With our parameters, a detection ratio angle of $\pi/24$ was proposed to improve the quantum noise behavior.
The sensitivity of a Sagnac interferometer,
which retains all the ET-LF Michelson parameters, can be achieved
as shown in Fig.\,\ref{fig:spe_compare} (blue curve).
A high-power Sagnac with, 180\,kW inside the arm cavities,
shows an improved sensitivity [Fig.\,\ref{fig:spe_compare} (red curve)]. This
gives a comparable sensitivity to the ET- LF Michelson interferometer
and should be considered here for its potential to further improve the
sensitivity in the long term.

A specific
homodyne detection can be precisely determined by using a
wave plate to shift the phase between the Michelson signal and Sagnac signal.
Additionally we have conducted a preliminary investigation into
the possible error signals which could be used to control a
realistic Sagnac interferometer with PBS leakage, including
looking at potential error signals generated from the leaked light
provided by the imperfect PBS.  Potential error signals for the
required degrees of freedom, PRCL, CARM, DARM, and MICH,
were generated using \textsc{Finesse}.  While the design of a
full control scheme for the proposed Sagnac configuration is beyond the
scope of this paper, our preliminary results are encouraging
and suggest that a control scheme can be developed by utilizing
the different interferometer responses of the light fields in the two
polarization states.


\section{Acknowledgements}
\label{acknowledgement}
This work has been supported by the Science and Technology Facilities Council
(STFC). We would like to thank Haixing Miao, Yanbei Chen and Stefan Hild for useful 
discussions and Harald L\"{u}ck, Stefan Danilishin, Farid Khalili, and Chiara Mingarelli 
for comments and suggestions. This document has been assigned the LIGO Laboratory 
Document No. LIGO-P1300035.


\appendix


\section{Optical System Block Diagram}\label{apps:BD}
\subsection{Lossless free space propagation and single mirror reflection}
	  \begin{figure}[!t]
	  \centering
 	   \includegraphics[width=0.45\textwidth,keepaspectratio]{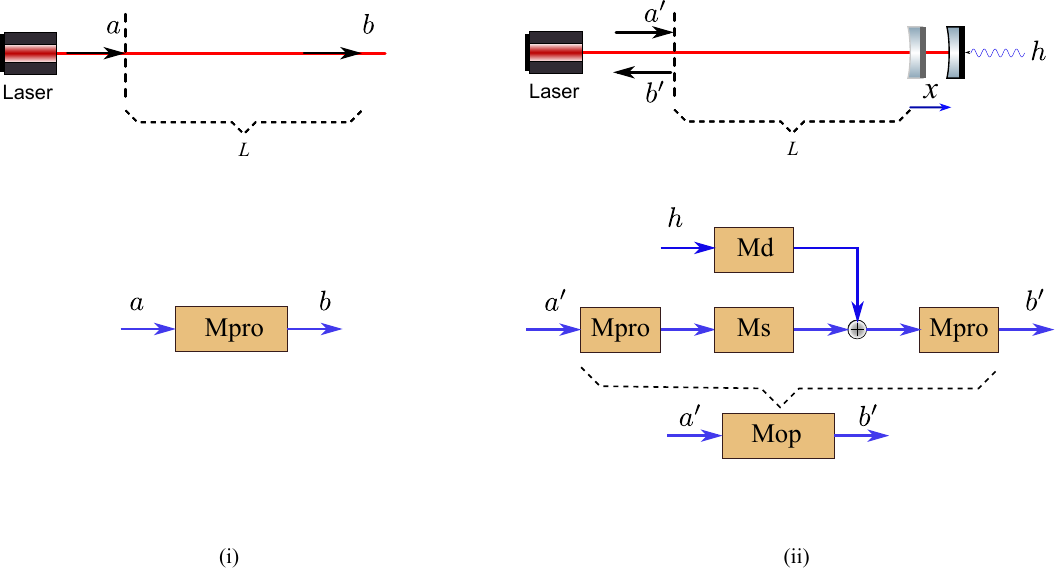}
    	   \caption{Diagrams showing the schematics and the corresponding
	   block diagrams of a light
	   field (\rmnum{1}) propagating in a free space and (\rmnum{2})
	   reflected by a single free hanging perfect mirror. $a$,
	   $b$ and $a'$, $b'$ are the input and output variables, respectively.
	   $\bf M_{pro}$ is the propagating input-output relation, $\bf M_m$ is
	   the TF of a perfect mirror including the mechanical
	   response and $\bf H_m$ is the GW signal TF,
	   which couples in via the mirror's motion. We denote
	   $\bf{M}_{op}$ as the combination of the TFs of propagations
	   and reflection. }
   	   \label{fig:BD_sing}
\end{figure}
A block diagram is convenient for the description of a system which consists
of several principal parts. Each of them has a well-defined TF
(or say \textit{input-output relation}), in particular with an existing
closed loop.
Each block in the diagram represents each individual subsystem
which is described by a specific TF.  Different
blocks are connected by arrows (including the signal flow
direction), specifying the relationships between each blocks.
Here we present the block diagrams of two simple optical systems:  a laser
beam (\rmnum{1})  propagating in a free space and (\rmnum{2}) reflected by
a signal free hanging  perfect mirror as shown in Fig.\,\ref{fig:BD_sing}.
The GW single and radiation pressure force both act on the
free mirror, inducing a displacement $x$. According to the propagation
equations of an electromagnetic field, we thus have the output fields
{\small
\begin{align}
&\left[\begin{array}{c}
    b_1\\
    b_2\end{array}\right] = \bf{M}_{pro}\cdot \left[\begin{array}{c}
    a_1\\
    a_2\end{array}\right]\label{equ:BDinoutfree1},\\
&\left[\begin{array}{c}
    b'_1\\
    b'_2\end{array}\right]= \bf{M}_{op}\cdot \left[\begin{array}{c}
    a'_1\\
    a'_2\end{array}\right]+ {\bf M}_{\rm pro} \cdot {\bf H_m} \cdot \rm h,\label{equ:BDinoutsing1}
\end{align}
}with
{\small
\begin{align}
\nonumber&{\bf M_{pro}}=e^{i\phi}\mathbf{R}[\varphi],\qquad{\bf M_m}=\left[\begin{array}{cc}
	                 1 & 0 \\
	        -\kappa & 1\end{array}\right],\\
\nonumber &{\bf M}_{\rm op} = \bf{M}_{pro}\cdot\bf{M}_{m}\cdot {\bf M}_{pro}, \\
\nonumber&{\bf H_m}=\left[\begin{array}{c}
                     0\\
                     \frac{\sqrt{2\kappa}}{h_{SQL}}\end{array}\right],\qquad\kappa = \frac{8I_0\omega_0}{mc^2\Omega^2}, \qquad h_{SQL}=\sqrt{\frac{8\hbar}{m\Omega^2L^2}},\\
             &\mathbf{R}[\varphi] = \left[ \begin{array}{cc}
\cos \varphi&  -\sin \varphi\\
\sin \varphi & \cos \varphi  \end{array} \right],\qquad
\phi = \frac{\Omega L}{c}, \qquad \varphi = \frac{\omega_0 L}{c}.\label{equ:cavM}
\end{align}
}

\subsection{Lossless optical cavity}
    \begin{figure}[b]\centering
    \includegraphics[width=0.45\textwidth,keepaspectratio]{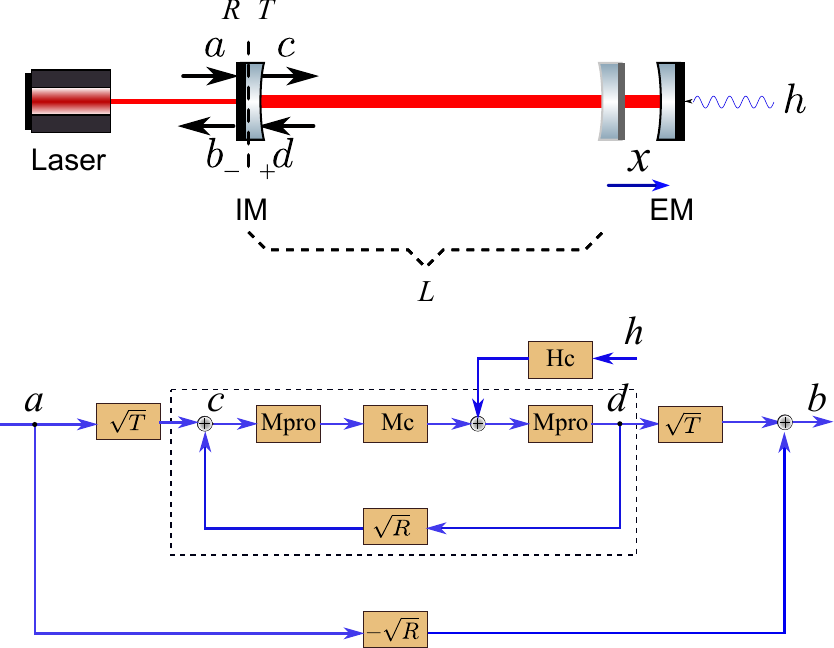}
    \caption{Diagram showing a schematic of an arbitrary lossless cavity and
    its block diagram. $a$,
    $b$, $c$ and $d$ represent
    light fields at different positions. $R$ and $T$ are the reflectivity and
    transmissivity of the cavity IM, which satisfy $R+T=1$ for the lossless
    case. The cavity EM is assumed to be perfect with a reflectivity equal to 1.
    A closed loop is formed due to the partial reflection of  the IM. The GW
    signal and
    radiation pressure force act on the EM only.}
    \label{fig:BD_cav}
    \end{figure}
With a similar process, we can describe an arbitrary
lossless cavity ($R+T=1$) with a partially transmissive input mirror (IM) and  a
perfect end mirror
(EM), as shown in Fig.\,\ref{fig:BD_cav}. Because of both the transmission and
reflection feature of the IM, a closed loop is formed (framed by the dashed
box in the diagram). Based on the general TF evaluation
of a system with a
feedback loop, the TF of the dashed box can be written as
{\small
\begin{equation}
\label{equ:mcav}{\bf M}_{\rm c} = \frac{1}{{\bf I}-\sqrt{R}\cdot \bf{M}_{\rm op}},
\end{equation}
}where ${\bf I}$ is the $2\times2$ \textit{identity matrix} and ${\bf M}_{op}$ is
defined in Eq.\,(\ref{equ:cavM}).

A cavity being either resonant or detuned can be schematically described by the
same diagram (see Fig.\,\ref{fig:BD_cav}). The only difference comes from the
propagation matrix $\rm M_{pro}$ due to different propagating lengths.
Consequently, we obtain the general expression of the output field $b$ of a lossless cavity as\footnote{Please be aware of the
order of the matrices. They follow the flow direction of the signal.}
{\small
\begin{align}
\nonumber \left[\begin{array}{c} {b}_1(\Omega)\\
	 {b}_2(\Omega)\end{array}\right] = &\left[-\sqrt{R} \cdot {\bf I}+T\cdot {\bf M}_{\rm c}  \cdot \bf{M}_{\rm op}\right] \left[\begin{array}{c} {a}_1(\Omega)\\
	 {a}_2(\Omega)\end{array}\right]\\
&+\left[\sqrt{T} \cdot {\bf M}_{\rm c} \cdot {\bf M}_{\rm pro} \cdot {\bf H}_{\rm c}\right]h(\Omega) \label{equ:BDinoutcavity}.
\end{align}
}where ${\bf M}_{\rm pro}$, ${\bf M}_{\rm op}$, and ${\bf H}_{\rm c}$ are evaluated
according to the variable definitions in Eq.\,(\ref{equ:cavM}). However, these
parameters need to be modified as the laser power incident on the EM
is different;  ${\bf M}_{\rm c}$ is defined in Eq.\,(\ref{equ:mcav}). For a resonant
cavity, the propagation matrix is
{\small
\begin{equation}
{\bf M}_{\rm pro}^{\rm re}= e^{i \phi}.
\end{equation}
}The output field of a resonant lossless cavity can then be expressed by
{\small
\begin{align}
\nonumber \left[\begin{array}{c} {b}_1(\Omega)\\
	 {b}_2(\Omega)\end{array}\right]
	\nonumber=&e^{2i\phi_{cav}}\left[ \begin{array}{cc}
1&  0\\
-\kappa_{cav} & 1  \end{array} \right] \left[\begin{array}{c} {a}_1(\Omega)\\
	 {a}_2(\Omega)\end{array}\right]\\
	&+e^{i\phi_{cav}}\frac{\sqrt{2\kappa_{cav}}}{h_{SQL}}\left[\begin{array}{c}0\\
	1\end{array}\right]h(\Omega)\label{equ:BDinoutcavity_ana},
\end{align}
}with
{\small
\begin{align}
\phi_{cav} &= \arctan{(\frac{1+\sqrt{R_i}}{1-\sqrt{R_i}}}\tan{\Omega\tau)},\\
\kappa_{cav}&=\frac{T_i \kappa}{1-2\sqrt{R_i}\cos(2\Omega\tau)+R_i}.\qquad
\end{align}
}By inserting the distinctive propagation matrix of a detuned cavity with detuning phase $\theta$
{\small
\begin{equation}
{\bf M}^{\rm de}_{\rm pro}=e^{i\phi} {\bf R}[\theta],
\end{equation}
}a similar output can be obtained. The outputs for both resonant and detuned cavities match with the results shown in \cite{phd.Miao}.

\subsection{Lossy optical cavity}
 \begin{figure}[b]\centering
    \includegraphics[width=0.45\textwidth,keepaspectratio]{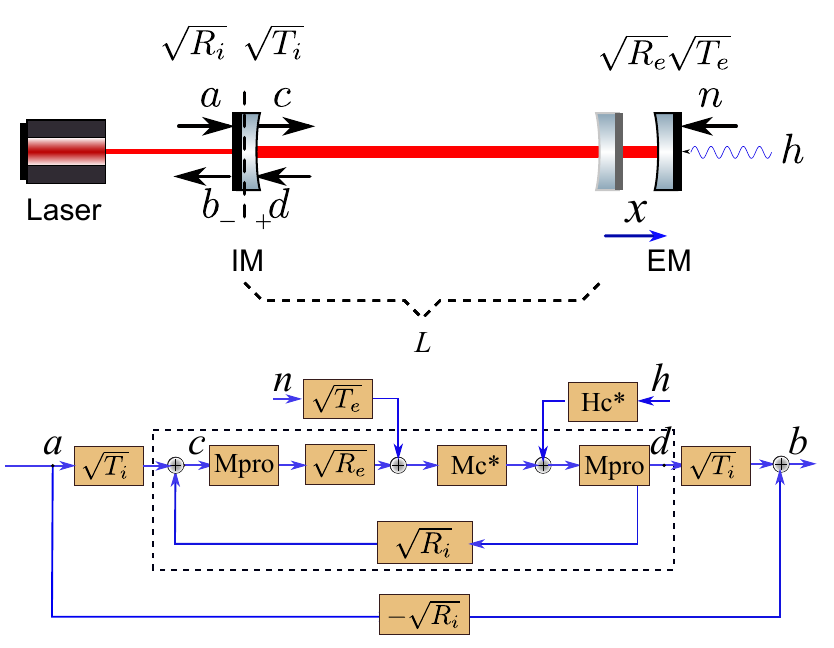}
    \caption{Diagram showing a schematic of a lossy cavity and its corresponding
    block diagram.
    Both the IM and EM are partial transmissive mirrors. Cavity losses are
    grouped into the transmission of the EM, namely laser beams transmitted by
    the EM
    are considered as optical losses. Vacuum fluctuations $n$ simultaneously
    couple in due to these optical losses.The reflectivity and transmissivity of the
    mirrors still satisfy equation $T_{i,e}+R_{i,e}=1$. A mechanical displacement
    $x$ due to the circulating light radiation pressure force occurs only at
    the EM. The GW signal acts on the EM only. }
    \label{fig:BD_arm_loss}
    \end{figure}
\begin{figure}[ht]\centering
    \includegraphics[width=0.25\textwidth,keepaspectratio]{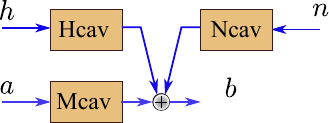}
    \caption{A block diagram of a general lossy resonant optical cavity. The
    TFs are defined in Eqs.\,(\ref{equ:BDinoutlossycavity_ana}) and
    (\ref{equ:BDinoutlossycavity_format}).}
    \label{fig:BD_arm_format}
    \end{figure}

We further consider the case of a cavity where optical losses have been included.
It has been shown by previous work~\cite{Kimble02, Chen03, Purdue02, Purdue02a} that
additional vacuum fluctuations simultaneously enter into the optical system at
different ports when there are optical losses. In our work, these optical loss
induced vacuum fluctuations are grouped into one port, the cavity EM, and
denoted as $n$. A similar schematic of a lossy cavity is shown in
Fig.\,\ref{fig:BD_arm_loss}. We thus can write down the input-output relation of a
lossy cavity as
{\small
\begin{align}
\nonumber \left[\begin{array}{c} {b}_1(\Omega)\\
	 {b}_2(\Omega)\end{array}\right] = &\left[-\sqrt{R_i} \cdot {\bf I}+T_i\sqrt{R_e}\cdot {\bf M}_{\rm cav}^*  \cdot \bf{M}_{\rm op}^*\right] \left[\begin{array}{c} {a}_1(\Omega)\\
	 {a}_2(\Omega)\end{array}\right]\\
\nonumber&+\left[\sqrt{T_i}\sqrt{R_e} \cdot {\bf M}_{\rm cav}^* \cdot {\bf M}_{\rm pro} \cdot {\bf H}_{\rm c}^*\right]h(\Omega) \label{equ:BDinoutlossycavity} ,\\
&+\left[\sqrt{T_e}\cdot {\bf M}_{\rm cav}^*\cdot {\bf M}_{\rm pro}\right] \left[\begin{array}{c} {n}_1(\Omega)\\
	 {n}_2(\Omega)\end{array}\right],
\end{align}
}with
{\small\[{\bf M}_{\rm cav}^*  = \frac{1}{{\bf I}-\sqrt{R_iR_e}\cdot \bf{M}_{\rm op}^*}.\]
}$\bf{M}_{c}^*$, $\bf{M}_{\rm op}^*$, $\bf{H}_{c}^*$ have the same
format as in the lossless expressions in Eqs.\,(\ref{equ:cavM}) and
(\ref{equ:mcav}),
but replacing $\kappa$ by
\small
\begin{align}
\label{equ:kappaloss}\kappa^*= \sqrt{R_e} \frac{8I^*\omega_0}{mc^2\Omega^2},
\end{align}\normalsize
due to the EM loss, resulting in a lower laser power circulating inside the cavity.
We assume the cavity loss is far smaller than 1, $T_e\ll1$. This enables an
approximation by keeping the leading order of $\sqrt{T_e}$ in the input-output relation. For a resonant lossy cavity, we thus get the same format input-output relation as Eq.\,(\ref{equ:BDinoutcavity_ana}):
{\small
\begin{align}
\nonumber \left[\begin{array}{c} {b}_1(\Omega)\\
	 {b}_2(\Omega)\end{array}\right] = &e^{2i\phi_{cav}}\left[ \begin{array}{cc}
1&  0\\
-\kappa_{cav}^* & 1  \end{array} \right] \left[\begin{array}{c} {a}_1(\Omega)\\
	 {a}_2(\Omega)\end{array}\right]\\
\nonumber	&+e^{i\phi_{cav}}\frac{\sqrt{2\kappa_{cav}^*}}{h_{SQL}}\left[\begin{array}{c}0\\
	1\end{array}\right]h(\Omega)\\
	&\sqrt{T_e}e^{i\phi_{cav}}\sqrt{\frac{\kappa_{cav}^*}{\kappa^*}}\left[ \begin{array}{cc}
1&  0\\-
\sqrt{\frac{\kappa_{cav}^*}{T_i}} e^{i\phi_{cav}-i\Omega\tau} & 1  \end{array} \right] \left[\begin{array}{c} {n}_1(\Omega)\\
	 {n}_2(\Omega)\end{array}\right]
	\label{equ:BDinoutlossycavity_ana}. \end{align}
}We turn this input-output relation into
{\small
\begin{equation}
 \left[\begin{array}{c} {b}_1(\Omega)\\
	 {b}_2(\Omega)\end{array}\right] = {\bf M}_{\rm cav} \left[\begin{array}{c} {a}_1(\Omega)\\
	 {a}_2(\Omega)\end{array}\right]
	+{\bf H}_{\rm cav}h(\Omega)+{\bf N}_{\rm cav}\left[\begin{array}{c} {n}_1(\Omega)\\
	 {n}_2(\Omega)\end{array}\right],
	\label{equ:BDinoutlossycavity_format}
	\end{equation}
}where ${\bf M}_{\rm cav}$, ${\bf H}_{\rm cav}$ and ${\bf N}_{\rm cav}$
correspond to the matrices in  Eq.\,(\ref{equ:BDinoutlossycavity_ana}). This gives
the general input-output relation expression of a lossy resonant cavity. We further
simplify such a cavity into a new block diagram as shown in
Fig.\,\ref{fig:BD_arm_format}, which can be recalled by any optical system containing a resonant cavity (i.e., interferometer arm cavities).

\section{Imperfect Polarizing beamsplitter}\label{apps:PBS}
  \begin{figure}[t]\centering
    \includegraphics[width=0.3\textwidth,keepaspectratio]{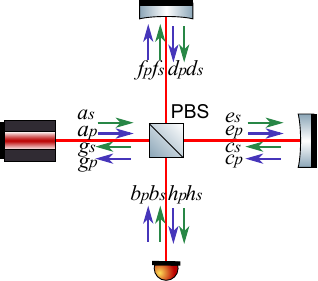}
    \caption{Figure showing the relations of polarized light fields at the four ports of
    a PBS.  Subscripts
    $s$ and $p$ represent the polarization of each field. The arrows denote the
    {\it p}-polarized (blue) and {\it s}-polarized (green) beams.
    The PBS has an extinction ratio of $\eta_p$ for the transmitted {\it p}-polarized
    beam and $\eta_s$ for the reflected {\it s}-polarized beam. }
    \label{fig:PBS}
    \end{figure}
Here we specify the relations of the transmission and reflection fields at a polarizing
beam splitter (PBS).
The PBS as shown in Fig.\,\ref{fig:PBS} has an extinction ratio of $\eta_p$ for
the transmitted {\it p}-polarized beam and $\eta_s$ for the reflected {\it s}-polarized
beam. Therefore, the relations between the input and output fields are
{\small
\begin{align}
\nonumber e_s=&\sqrt{\eta_s}\cdot a_s+\sqrt{1-\eta_s}\cdot d_s, \\
\nonumber e_p=&\sqrt{\eta_p}\cdot d_p+\sqrt{1-\eta_p}\cdot a_p, \\
\nonumber f_s=&\sqrt{\eta_s}\cdot b_s+\sqrt{1-\eta_s}\cdot c_s, \\
\nonumber f_p=&\sqrt{\eta_p}\cdot c_p+\sqrt{1-\eta_p}\cdot b_p, \\
\nonumber g_s=&\sqrt{\eta_s}\cdot c_s+\sqrt{1-\eta_s}\cdot b_s, \\
\nonumber g_p=&\sqrt{\eta_p}\cdot b_p+\sqrt{1-\eta_p}\cdot c_p, \\
\nonumber h_s=&\sqrt{\eta_s}\cdot d_s+\sqrt{1-\eta_s}\cdot a_s, \\
\nonumber h_p=&\sqrt{\eta_p}\cdot a_p+\sqrt{1-\eta_p}\cdot d_p.
\end{align}
}The reflected {\it p}-polarized beams and transmitted {\it s}-polarized beams are also called {\it leakages}.

\section{Noise spectral density}\label{app:NSD}
With a well-defined input-output relation of an optical system, we can write
the quadrature equation in a general form as
{\small
\begin{align}
\nonumber\left[\begin{array}{c} {b}_1(\Omega)\\
	 {b}_2(\Omega)\end{array}\right] = &\left[\begin{array}{cc}
	                 M_{11} & M_{12} \\
	                 M_{21} & M_{22}\end{array}\right]\left[\begin{array}{c} {a}_1(\Omega)\\
	{a}_2(\Omega)\end{array}\right]+\left[\begin{array}{c}
                     D_1\\
	                 D_2\end{array}\right]h(\Omega)\\
&+\left[\begin{array}{cc}
	                 N_{11} & N_{12} \\
	                 N_{21} & N_{22}\end{array}\right]\left[\begin{array}{c} {n}_1(\Omega)\\
	 {n}_2(\Omega)\end{array}\right],
	\label{equ:generalinout}
\end{align}
}where each transfer matrix element is determined by a specific optical
layout, with the first item being the vacuum fluctuations of the light field (i.e., laser beam), the second item being the GW signal and the third one the vacuum fluctuation induced by optical losses. Homodyne detection offers a possible selection of the readout as
{\small
\begin{equation}
 {b}_{\zeta}(\Omega) =  {b}_1(\Omega)\cos\zeta + {b}_2(\Omega)\sin\zeta,
\label{equ:homodyneGen}
\end{equation}
}with the \textit{homodyne detection angle} $\zeta$.
Correspondingly, the GW strain $h$-normalized quantum NSD of this
measurement is defined as
{\small
\begin{align}
S_h(\Omega)=\frac{(\cos\zeta,\sin\zeta){\bf M \cdot (S_a+S_n)\cdot M}^{\rm \dag}(\cos\zeta,\sin\zeta)^T}{|D_1\cos\xi+D_2\sin\xi|^2}, 
\label{equ:noisespe}
\end{align}
}where $n$ is the optical loss induced vacuum fluctuations. The NSD matrix
${\bf S_n}$ thus is always an identity matrix. ${\bf S_{a}}$ is NSD matrix induced
by the laser  fluctuations and the elements are the corresponding NSD defined in \cite{Kimble02} as
{\small
\begin{equation}
{\bf S_a}(\Omega)=\left[\begin{array}{cc}
       S_{a_1a_1}(\Omega) & S_{a_1a_2}(\Omega)\\
       S_{a_2a_1}(\Omega) & S_{a_2a_2}(\Omega)\end{array}\right],
\end{equation}
}with $\pi S_{a_ia_j}(\Omega)\delta(\Omega-\Omega')=\langle \rm{in}|
 {a}_i(\Omega) {a}^{\dag}_{j}(\Omega ')|\rm{in}\rangle_{sym}$, i.e., if $|\rm{in}\rangle$ is a vacuum input $|0\rangle$, then $S_{a_1a_1}(\Omega) =S_{a_2a_2}(\Omega)=1$ and $S_{a_1a_2}(\Omega)=S_{a_2a_1}(\Omega)=0 $ lead to an identity matrix ${\bf S}_a={\bf S}_{\rm vac}={\bf I}$.
 \\
 \section{Complete equations}\label{app:full}
Here we present the complete equations of the two polarized output fields from a
Sagnac interferometer with an imperfect PBS [see Eqs.\,\eqref{equ:inout_p} and
\eqref{equ:inout_s}].
{\small\begin{align}
\nonumber \left[\begin{array}{c} {q}_{p1}(\Omega)\\
	 {q}_{p2}(\Omega)\end{array}\right] =
	&\left[-\sqrt{\eta_p}{\bf I}+(1-\eta_p)\sqrt{1-\eta_s}
	{\bf M}_{\rm CLG}{\bf M}_{\rm sag}\right]
	\left[\begin{array}{c} {g}_{p1}(\Omega)\\
	 {g}_{p2}(\Omega)\end{array}\right] \\
\nonumber&+\sqrt{1-\eta_p}\sqrt{\eta_s}{\bf M}_{\rm CLG}{\bf M}_{\rm arm}
         \left[\begin{array}{c} {g}_{s1}(\Omega)\\
	 {g}_{s2}(\Omega)\end{array}\right]\\
\nonumber&+\sqrt{1-\eta_p}\sqrt{1-\eta_s}
         {\bf M}_{\rm CLG}{\bf H}_{\rm sag}h(\Omega)\\
\label{equ:inout_p1}&+\sqrt{1-\eta_p}\sqrt{1-\eta_s}
         {\bf M}_{\rm CLG}{\bf N}_{\rm sag}
         \left[\begin{array}{c} {n}_{1}(\Omega)\\
	 {n}_{2}(\Omega)\end{array}\right]\,,\\
\nonumber\left[\begin{array}{c} {q}_{s1}(\Omega)\\
	 {q}_{s2}(\Omega)\end{array}\right] =
	&-\left[-\sqrt{1-\eta_s}{\bf I}+\eta_s\sqrt{\eta_p}{\bf M}_{\rm CLG}
	{\bf M}_{\rm sag}\right]
	\left[\begin{array}{c} {g}_{s1}(\Omega)\\
	 {g}_{s2}(\Omega)\end{array}\right]\\
\nonumber&+\sqrt{\eta_s}\sqrt{1-\eta_p}{\bf M}_{\rm CLG}{\bf M}_{\rm arm}
	\left[\begin{array}{c} {g}_{p1}(\Omega)\\
	 {g}_{p2}(\Omega)\end{array}\right]\\
\nonumber&+\sqrt{\eta_s}{\bf M}_{\rm CLG}{\bf H}_{\rm arm}h(\Omega)\\
\label{equ:inout_s1}&+\sqrt{\eta_s}{\bf M}_{\rm CLG}{\bf N}_{\rm arm}
   	\left[\begin{array}{c} {n}_{1}(\Omega)\\
	 {n}_{2}(\Omega)\end{array}\right].
\end{align}
}


\bibliography{sagnac}

\end{document}